\documentclass[conference,compsoc]{IEEEtran}

\usepackage{amsmath}
\usepackage{amsthm}
\usepackage{xcolor}
\usepackage{csquotes}
\usepackage{balance}
\usepackage{sidecap}
\usepackage{multirow}
\usepackage{subcaption}
\usepackage{float}
\usepackage[numbers]{natbib}
\newcommand{\sys}{{DREAM}~}
\usepackage{enumitem}
\usepackage{tikz}
\usepackage{makecell} 
\usepackage{pifont}
\usepackage{hyperref}
\usepackage[nocompress]{cite}
\usepackage{fontawesome5}  %
\usepackage{pifont}  %
\usepackage{xcolor}  %
\usepackage{booktabs}
\usepackage{xurl}
\usepackage{wasysym}
\hypersetup{
    colorlinks=true,
    linkcolor=red,
    citecolor=green,
    filecolor=black,
    pdfstartview=Fit,
    breaklinks=true
}

\usepackage{ifpdf}

\newtheorem{definition}{Definition}
\newtheorem{theorem}{Theorem}

\usepackage{xcolor}
\usepackage{tcolorbox}

\usepackage{amssymb}
\usepackage{algorithm}
\usepackage{algcompatible}
\newtcolorbox{takeawaybox}{
  colback=gray!20,
  colframe=gray!20,
  coltitle=black,
  arc=4pt,
  boxrule=0.5pt,
  boxsep=2pt,
  left=2pt,
  right=2pt,
  top=2pt,
  bottom=2pt,
  before skip=0.7\baselineskip,
  after skip=0.7\baselineskip
}
\usepackage{algorithmicx}

\usepackage[table]{xcolor}

\ifCLASSOPTIONcompsoc
  \usepackage[nocompress]{cite}
\else
  \usepackage{cite}
\fi

\ifCLASSINFOpdf
\else
\fi

\begin{document}
\title{DREAM: Scalable Red Teaming for Text-to-Image Generative Systems\\ via Distribution Modeling}

\author{
    \IEEEauthorblockN{
        Boheng Li\textsuperscript{1}, 
        Junjie Wang\textsuperscript{2}, 
        Yiming Li\textsuperscript{1}\textsuperscript{*}, 
        Zhiyang Hu\textsuperscript{2}, 
        Leyi Qi\textsuperscript{3}, \\
        Jianshuo Dong\textsuperscript{4}, 
        Run Wang\textsuperscript{2}, 
        Han Qiu\textsuperscript{4}, 
        Zhan Qin\textsuperscript{3}, 
        Tianwei Zhang\textsuperscript{1}
    }
    \IEEEauthorblockA{\textsuperscript{1}Nanyang Technological University, boheng001@e.ntu.edu.sg, \{ym.li, tianwei.zhang\}@ntu.edu.sg}
    \IEEEauthorblockA{\textsuperscript{2}School of Cyber Science and Engineering, Wuhan University, \{junjie.wang, zhiyanghu, wangrun\}@whu.edu.cn}
    \IEEEauthorblockA{\textsuperscript{3}State Key Laboratory of Blockchain and Data Security, Zhejiang University, qileyi@foxmail.com, qinzhan@zju.edu.cn}
    \IEEEauthorblockA{\textsuperscript{4}Tsinghua University, dongjs23@mails.tsinghua.edu.cn, qiuhan@tsinghua.edu.cn}
    \IEEEauthorblockA{\textsuperscript{*}Corresponding Author}
}

\maketitle

\IEEEpeerreviewmaketitle

\begin{abstract}
Despite the integration of safety alignment and external filters, text-to-image (T2I) generative systems are still susceptible to producing harmful content, such as sexual or violent imagery. This raises serious concerns about unintended exposure and potential misuse. Red teaming, which aims to proactively identify diverse prompts that can elicit unsafe outputs from the T2I system, is increasingly recognized as an essential method for assessing and improving safety before real-world deployment. However, existing automated red teaming approaches often treat prompt discovery as an isolated, prompt-level optimization task, which limits their scalability, diversity, and overall effectiveness. To bridge this gap, in this paper, we propose DREAM, a scalable red teaming framework to automatically uncover diverse problematic prompts from a given T2I system. Unlike prior work that optimizes prompts individually, DREAM directly models the probabilistic distribution of the target system's problematic prompts, which enables explicit optimization over both effectiveness and diversity, and allows efficient large-scale sampling after training. To achieve this without direct access to representative training samples, we draw inspiration from energy-based models and reformulate the objective into a simple and tractable form. We further introduce GC-SPSA, an efficient optimization algorithm that provides stable gradient estimates through the long and potentially non-differentiable T2I pipeline. During inference, we also propose a diversity-aware sampling strategy to enhance prompt variety. The effectiveness of DREAM is validated through extensive experiments, demonstrating state-of-the-art performance across a wide range of T2I models and safety filters in terms of both prompt success rate and diversity. Our code is available at \href{https://github.com/AntigoneRandy/DREAM}{https://github.com/AntigoneRandy/DREAM}.
\end{abstract}

\section{Introduction}
Text-to-image (T2I) generative models \citep{croitoru2023diffusion,yang2023diffusion,rombach2022ldm,podell2024sdxl} are driving a new wave of visual content creation, reshaping our expectations of what machines are capable of. Trained on large-scale datasets \citep{schuhmann2022laion}, these models capture rich associations between language and imagery, allowing them to produce high-quality images with simple text inputs (known as \emph{prompts}). Their ease of use and impressive flexibility have driven rapid adoption across creative arts, entertainment, and social media, particularly among younger users such as teenagers \citep{wang2023factors,commonsense2024teens,ali2023constructing}. However, the same large-scale, web-crawled datasets that enable this versatility also inevitably contain not safe for work (NSFW) content (e.g., sexually explicit material) \citep{schramowski2023sld,gandikota2023erasing}. As a result, the models also acquire the ability to produce harmful images during real use, raising serious ethical, legal, and accountability concerns \citep{qu2023unsafe,ho2023development}.

To mitigate these risks, a growing number of efforts from both academia and industry~\citep{kumari2023ablating,gandikota2023erasing,gandikota2024unified,compvis2022sc,nsfw_text_classifier} have focused on improving the safety of T2I generative models. One popular approach is \emph{safety alignment}, also referred to as \emph{unsafe concept erasure} in the T2I literature \citep{kumari2023ablating,gandikota2023erasing,gandikota2024unified,gong2024rece}, which fine-tunes the model using a curated set of unsafe prompts or images to suppress undesirable generations. This process helps steer the model toward harmless outputs: for example, returning a clothed figure even when prompted with ``a nude person''. In addition, commercial companies like Stability AI \citep{stabilityai2025} and Ideogram \citep{ideogram2025} also employ proprietary \emph{safety filters} (e.g., NSFW image detectors) to block generation attempts when unsafe content is detected. These filters, when combined with the core generative model and other processing components, constitute the deployed \emph{T2I system}. However, while these techniques show promising results in controlled environments, they remain imperfect when applied in practice. For example, both real-world users and researchers \citep{chin2024p4d,li2024art} have reported that prompts unseen during training (e.g., implicit references to sensitive content), or even totally benign inputs (e.g., ``the origin of woman''), may still escape moderation and lead to unsafe outputs. These observations highlight the limitations of current methods under open-ended inputs and the urgent need for proactive mechanisms to expose safety vulnerabilities {of T2I generative systems} before real-world deployment. 

One emerging solution to proactively identify such blind spots is \textit{red teaming}, where model owners (e.g., developers) simulate the behavior of real-world users to generate various testing prompts, aiming to systematically probe the model's failure modes before deployment. In the context of T2I generative systems, red teaming typically attempts to find a diverse set of problematic prompts that can elicit unsafe or policy-violating outputs despite potential safeguards \citep{chin2024p4d,li2024art}. By doing so, it not only serves as an evaluation tool for stress-testing the system's safety and trustworthiness under open-ended inputs \citep{chin2024p4d}, but also provides valuable references for future improvement \citep{liu2024safetydpo}. As a result, red teaming is increasingly recognized as a critical practice, with major companies like Google \citep{quaye2024advnibbler} initiating human-in-the-loop red teaming programs. At the same time, regulatory bodies are increasingly emphasizing rigorous safety testing before deployment, as reflected in the EU AI Act \citep{euai2024} and the U.S. NIST AI Risk Management Framework \citep{nist2023}, alongside similar official efforts around the world~\citep{uk2024,singapore2024}.

While early red teaming relied on human experts, recent works~\citep{chin2024p4d,li2024art,mehrabi2024flirt} have shifted toward automated red teaming, aiming to discover problematic prompts with minimal human intervention. For example, FLIRT~\citep{mehrabi2024flirt} employs a large language model (LLM) to iteratively rewrite a seed prompt toward unsafe outputs, while P4D~\citep{chin2024p4d} starts with a moderated unsafe prompt and applies token-level gradient-based substitutions to penetrate safety alignment. However, these methods often struggle to balance the success rate with prompt diversity, and can be prohibitively slow and costly to scale. These limitations underscore the urgent need for a scalable red teaming method that can efficiently generate a large, diverse set of effective problematic prompts.

In this paper, we present the first attempt to bridge the aforementioned gap. Our method is driven by a unified insight into the shared limitations of prior works: they treat red teaming as a prompt-to-prompt discrete optimization problem, where each seed prompt is optimized independently yet without accumulating global, distribution-level knowledge of the target model's unsafe prompt distribution across runs. Built upon this understanding, we propose \textbf{\underline{D}}istributional \textbf{\underline{R}}ed t\textbf{\underline{EA}}ming via energy-based \textbf{\underline{M}}odeling (DREAM), which directly models the probabilistic distribution of the target model's unsafe prompts by training a parameterized prompt generator (e.g., an autoregressive LLM).  In contrast to previous approaches, our formulation enables explicit optimization of both success and diversity, supports global updates to the modeled prompt distribution, and allows efficient large-scale sampling after training. 

However, modeling the target prompt distribution is challenging, as it is tightly coupled to the specific T2I system and lacks sufficiently representative samples, making direct training infeasible. To overcome this, we draw inspiration from energy-based models \citep{lecun2006tutorial} and decompose the originally intractable training objective into two simple sub-objectives that allow effective distribution learning without direct sample access. Moreover, to enable effective and efficient gradient-based optimization for these objectives under long and potentially non-differentiable pipelines, we introduce Gradient-Calibrated Simultaneous Perturbation Stochastic Approximation (GC-SPSA), an efficient zeroth-order optimization method based on SPSA \citep{spall1992multivariate}. Specifically, it estimates gradients using only forward evaluations and further improves stability via a history-aware calibration mechanism. We also provide theoretical analysis and convergence guarantees to support GC-SPSA. Finally, we propose a novel adaptive temperature scaling method to further enhance coverage at inference time.

We conduct extensive experiments on 5 state-of-the-art (SOTA) safety-aligned T2I models and 4 safety filters, comparing \sys with 10 SOTA baselines across two unsafe categories. The results demonstrate that DREAM achieves superior performance in terms of prompt success rate while maintaining diversity comparable to human-written datasets. \sys also generalizes well to advanced T2I models (e.g., SDXL, SD v3) and 6 other NSFW themes, and remains effective even under combinational defenses or aggressive filters. We further conduct a user study to confirm that the red-teaming prompts generated by DREAM are both diverse and highly effective from human perspectives. Moreover, case studies demonstrate that the prompts generated by DREAM transfer effectively to 4 commercial T2I platforms with unknown safety mechanisms. Finally, DREAM also enhances safety fine-tuning, enabling defended models to better resist both seen and unseen harmful prompts.

To summarize, we make the following key contributions:
\begin{itemize}[leftmargin=*]
    \item We revisit existing red teaming methods and identify a shared limitation: they treat prompt discovery as an isolated, prompt-level optimization problem without global modeling, restricting their scalability and performance.
    \item We introduce DREAM, a scalable and distribution-aware red teaming framework that learns a probabilistic model over unsafe prompts using energy-based modeling. We further propose GC-SPSA, a novel zeroth-order optimizer that supports effective and efficient training, along with adaptive inference strategies for broader coverage. We also provide theoretical analyses and global convergence guarantees to support our GC-SPSA framework.
    \item We conduct comprehensive evaluations across 5 safety-aligned T2I models, 4 safety filters, and 10 SOTA baselines, showing that \sys achieves superior prompt success rates while matching human-level diversity. We also show that DREAM can expose failure cases in 4 commercial T2I platforms and improve safety fine-tuning with notable generalization to unseen harmful prompts.
\end{itemize}

\section{Related Work}
\subsection{Text-to-Image Generative Models}
Text-to-image (T2I) generative models have become a cornerstone of modern visual synthesis, enabling users to create highly detailed images from natural language descriptions. Among various generative paradigms, diffusion models \citep{ho2020ddpm,rombach2022ldm, croitoru2023diffusion, yang2023diffusion, podell2024sdxl} have emerged as the dominant approach due to their superior training stability, generation quality, and controllability. Diffusion models operate by iteratively denoising random noise into coherent images, often conditioned on texts, making them particularly effective for large-scale training and text-controlled generation. Building upon this, a wide range of open-source (e.g., Stable Diffusion family \citep{podell2024sdxl}) and commercial systems (e.g., DeepAI \citep{deepai2025}, DALL·E 3 \citep{openai_dalle3_2023}, Midjourney \citep{midjourney2025}, Ideogram \citep{ideogram2025}) have been developed, providing state-of-the-art generation experiences through user-friendly graphical interfaces.

\subsection{Unsafe Generation \& Mitigation}
The success of modern T2I models relies heavily on large datasets. For instance, Stable Diffusion is trained on LAION-5B \citep{schuhmann2022laion}, a web-scraped set of over 5 billion image-text pairs, while commercial models like Ideogram use even larger private datasets \citep{ideogram2025}. These datasets support powerful multimodal learning but also contain harmful content such as sexual or violent imagery. Such content can be reproduced by these models, raising ethical and legal concerns. The risks become particularly significant as these tools grow more accessible and popular among children and adolescents, who may experience psychological harm, safety risks, and developmental disruptions due to exposure \citep{qu2023unsafe,schramowski2023sld,guo2024moderating}.

In response to these concerns, several mitigation strategies have emerged, which can be broadly categorized into two lines: model safety alignment and inference-time safety filtering. Model safety alignment \citep{gandikota2023erasing,gandikota2024unified} refers to techniques that tune the diffusion model's parameters directly to suppress its ability to produce unsafe content. This is typically achieved by collecting a curated set of harmful prompts or images and reinforcing the model to ``unlearn'' them through methods such as adversarial training \citep{zhang2024defensive}, supervised fine-tuning \citep{gandikota2023erasing,kumari2023ablating}, or model editing \citep{gandikota2024unified,gong2024rece}. For example, CA~\citep{kumari2023ablating} fine-tunes the diffusion model to match the image distribution of an unsafe \emph{target concept} (e.g., ``a nude person'') to that of a safe \emph{anchor concept} (e.g., ``person''). As a result, the model learns to resist prompts that are the same or similar to training-time target concepts and generates safe images instead. In contrast, safety filters \citep{nsfw_text_classifier,chhabra_nsfw_detection_dl,compvis2022sc} act as external control mechanisms during inference. They can operate at prompt-level or image-level, aiming to detect and block unsafe content before or after generation. A representative case is the Safety Checker (SC) \citep{compvis2022sc} employed in Stable Diffusion models, which compares the generated image with a set of predefined \emph{sensitive concepts} and blocks outputs that exceed a similarity threshold.

While these approaches have demonstrated effectiveness with acceptable trade-offs in benign performance under their own evaluation protocols, their robustness in real-world scenarios has been frequently challenged by a growing body of recent research and user reports. For instance, text-based safety filters can be bypassed using simple synonym substitutions \citep{ba2024surrogateprompt}, while image-based filters may lose effectiveness under subtle alterations in image styles, compositions, or rendering \citep{reddit_dalle3_jailbreak}. Moreover, while safety-aligned models perform well when the prompts contain in-distribution explicit words, they may still fail to handle out-of-distribution veiled expressions, metaphors, or context-related implications \citep{li2024safegen,meng2025concept,chin2024p4d}, which are unseen during unlearning. In addition to these scattered findings, recent research \citep{yang2024sneakyprompt,yang2024mma,huang2025perception,zhuang2023qfattack} has developed various optimization methods to transform a given rejected unsafe prompt into evasive variants to bypass safety mechanisms, a technique known as \emph{adversarial jailbreak attacks}. These diverse failure patterns across different safety mechanisms suggest it is crucial to proactively test and improve the T2I generative system's safety before real-world deployment.

\subsection{Red Teaming for Text-to-Image Models}

The concept of ``red teaming'' originated during the Cold War era in the 1960s as a form of structured military system testing and has since expanded to fields like cybersecurity, airport security, software engineering, and more recently to AI and ML systems \citep{feffer2024red}.  For generative models, red teaming typically involves simulating real user behavior to explore the system and find prompts that produce harmful or policy-violating outputs \citep{verma2024operationalizing,hong2024curiosity,perez2022red}. Unlike jailbreak attacks that tweak known unsafe prompts into evasive variants \citep{yang2024sneakyprompt,huang2025perception}, red teaming focuses on broader exploration to reveal diverse or even unexpected failure modes \citep{li2024art}. It is now a key part of responsible model development \citep{quaye2024advnibbler} and is increasingly emphasized in recent regulatory frameworks \citep{euai2024,nist2023,uk2024}.

One predominant form of red teaming is manual construction. For example, the I2P dataset~\citep{yang2024sneakyprompt} was formed by collecting and filtering harmful prompts from various forums through a mix of automatic tools and human curation. Similarly, Google's Adversarial Nibbler Challenge \citep{quaye2024advnibbler} invited participants to attack real-world T2I models and selected high-quality prompts based on their effectiveness and diversity. Commercial providers also employ in-house or external experts to manually test models for discovering failure modes~\citep{openaiRedTeaming}. While such methods can surface unexpected and model-specific vulnerabilities, they rely heavily on human effort and lack automation, making them inefficient and expensive to conduct.

To this end, several methods for automated red teaming have been proposed \citep{chin2024p4d,mehrabi2024flirt,li2024art,tsai2024ring}. These methods typically adopt paradigms and techniques similar to jailbreak attacks and transform a set of initial prompts into harmful ones, using methods like token-level substitution \citep{chin2024p4d,tsai2024ring} and LLM-rewrite \citep{mehrabi2024flirt,li2024art}. However, as we will identify in the following section, this inherited formulation inherently limits their effectiveness, exploration space, and efficiency, making them suboptimal for scalable red teaming.

\section{Preliminaries}

\subsection{Threat Model}

We consider the red team to be a benign (non-malicious) model owner aiming to proactively identify safety vulnerabilities in their own T2I generative system. Specifically, their goal is to find a set of diverse and effective prompts that can elicit unsafe or policy-violating outputs, in order to assess and improve safety before real-world use. Specifically, we assume the red teamer (1) has full control over their T2I generative system. They may request it with an arbitrary prompt and receive the resulting image (or an all-black image if blocked by filters), or access the model's parameters and gradients. Note that our DREAM only requires query access to the target model, making it applicable to both white-box and black-box systems.  (2) can leverage auxiliary models (e.g., open-source LLMs) for assistance and has moderate computational resources to fine-tune these models.

\subsection{Formulation of Red Teaming}
Despite the growing importance of red teaming in evaluating the safety of T2I generative models, existing literature \citep{li2024art,mehrabi2024flirt,chin2024p4d} largely lacks a formal formulation that captures the fundamental nature of the red teaming task.

To bridge this gap, we present a formal definition of red teaming in this section. Let $\mathcal{X}$ and $\mathcal{Y}$ be the prompt space and image space of the target T2I generative system, respectively. We can draw the following definition:
\begin{definition}[Red Teaming T2I Systems]
\label{def:redteaming}
Let \( G: \mathcal{X} \rightarrow \mathcal{Y} \) be a T2I system that maps a text prompt \( x \in \mathcal{X} = \mathcal{V}^T \) to an image \( y \in \mathcal{Y} \), where \(\mathcal{V}\) and \(T\) represent the full vocabulary and the maximum prompt length of the system, respectively. Red teaming aims to find a prompt subset \(\mathcal{A}\subseteq\mathcal{X}\) such that:
\[
\mathcal{A} := \{x \in \mathcal{X} \mid \mathcal{O}(G(x)) = 1\},
\]
where \( \mathcal{O}: \mathcal{Y} \rightarrow \{0,1\} \) is a binary oracle classifier that outputs \(1\) if the image is unsafe, and \(0\) if the image is safe or the request is blocked by the built-in safety filter.
\end{definition}
Intuitively, this definition formulates red teaming as a combinatorial subset discovery problem that aims to identify all prompts within the full prompt set $\mathcal{V}^T$ that can trigger the T2I system to produce unsafe content.  Note that the oracle function \( \mathcal{O} \) is fundamentally unobservable in practice, as determining whether an image is ``unsafe'' is often vague, influenced by context, culture, and subjective interpretation \citep{schramowski2023sld}. As a practical alternative, red teaming methods rely on a \textit{surrogate scoring function} \( S: \mathcal{Y} \rightarrow \mathbb{R} \), which approximates the oracle with an objective score (e.g., the confidence score of an NSFW image detector). A prompt is deemed unsafe if its surrogate score is large enough to exceed a threshold \( \tau \), yielding the surrogate unsafe set $\mathcal{A}_\tau := \{ x \in \mathcal{X} \mid S(G(x)) \geq \tau \}$. While the surrogate formulation makes the task operational, obtaining the exact solution of the unsafe set \(\mathcal{A}_\tau\) remains computationally intractable, as the task essentially reduces to a combinatorial search problem over the prompt space \(\mathcal{X}\), whose size grows exponentially with the prompt length \(T\), i.e., \( |\mathcal{X}| = |\mathcal{V}|^T \). In such combinatorial settings, exhaustive enumeration is the only general procedure that can ensure complete accuracy \citep{hartmanis1982computers}, yet it requires evaluating the surrogate score \(S(G(x))\) for every enumerated \(x \in \mathcal{V}^T\), making it computationally infeasible even for modest values of \(T\). As a result, exact discovery is impractical except in trivial cases.

Fortunately, previous works have shown that exact recovery of \(\mathcal{A}_\tau\) is often unnecessary. For instance, unlearning a moderate number of diverse and representative unsafe prompts is often sufficient to invalidate a much broader class of similar unsafe prompts \citep{gandikota2023erasing,gandikota2024unified,ribeiro2018semantically,pei2017deepxplore}. Consequently, the practical goal of red teaming shifts from full enumeration to the discovery of a representative and diverse subset \(\hat{\mathcal{A}} \subseteq \mathcal{A}_\tau\), which captures a wide range of unsafe prompts while remaining computationally tractable to obtain.

\begin{algorithm}[t]
\caption{A Generic Form of Existing Methods}
\label{alg:local_search}
\small
\begin{algorithmic}[1]
\Statex \textbf{Input:} Seed distribution $\pi(x)$, number of prompts $N$, max steps $T$, scoring function $S(\cdot)$, target T2I system $G(\cdot)$, threshold $\tau$, update operator $\textsc{Update}(\cdot)$
\Statex \textbf{Output:} Final set of optimized prompts $\hat{\mathcal{A}}$
\State $\hat{\mathcal{A}} \leftarrow \emptyset$

\FOR{$i = 1$ to $N$}
    \State $x_i^{(0)} \sim \pi(x)$
    \State $t \leftarrow 0$
    
    \WHILE{$t < T$ \textbf{and} $S(G(x_i^{(t)})) < \tau$}
        \State $x_i^{(t+1)} \leftarrow \textsc{Update}(x_i^{(t)}, S(G(x_i^{(t)})))$
        \State $t \leftarrow t + 1$
    \ENDWHILE

    \State $\hat{\mathcal{A}} \leftarrow \hat{\mathcal{A}} \cup \{x_i^{(t)}\}$
\ENDFOR

\State \textbf{Return} $\hat{\mathcal{A}}$
\end{algorithmic}
\end{algorithm}

\subsection{Limitations of Previous Works}
\label{sec:lim}
With an understanding of the red teaming task formulation, we now take a closer look at existing methods \citep{chin2024p4d,tsai2024ring,mehrabi2024flirt,li2024art}. While prior works differ substantially in their technical implementation, we distilled them into a unified, generic prompt-level discrete optimization paradigm, formalized in Alg.~\ref{alg:local_search}. Under this view, a red teaming algorithm begins with a seed prompt sampled from a seed distribution \(\pi(x)\), and iteratively applies an \textsc{Update} operator guided by a scoring function \(S(G(x))\), where \(G(x)\) denotes the image generated by the T2I model. This loop continues until a generation crosses a threshold or a step budget is reached, at which point the final prompt is collected and the process resets. Note that the seed prompt distribution, the \textsc{Update} operator, and the scoring function are all method-dependent. Despite empirical progress in uncovering unsafe prompts, we identify that this core algorithmic structure introduces two fundamental limitations, making these methods less suitable for scalable red teaming.

First, the \textsc{Update} operator essentially performs discrete optimization, which is inherently difficult due to the discontinuous and non-smooth nature of the ill-posed loss landscape of the discrete prompt space \citep{bengio2013estimating}. In fact, how to accurately obtain prompt-level gradient remains an open challenge in existing literature \citep{wen2023hard,wang2021adversarial}. As such, some existing methods \citep{yang2024sneakyprompt,yang2024mma} resort to token-level gradient replacement, where each token is iteratively and greedily updated based on its locally estimated gradient with respect to \(S(G(x_{i}^{(t)}))\). However, this limits the search space to local neighborhoods around the initial seed, making the optimization process highly sensitive to initialization \citep{jiang2025jailbreaking}.  While recent methods attempt to broaden the search space by prompting LLMs to generate sentence-level paraphrases~\citep{li2024art,mehrabi2024flirt}, these approaches are largely heuristic, lack convergence guarantees, and often result in unstable training dynamics in practice.

Second, one can easily observe from Alg.~\ref{alg:local_search} that current methods are essentially operating at the individual prompt level: each run starts from a fresh seed prompt \(x_i^{(0)}\), performs a local search trajectory \(\{x_i^{(1)}, x_i^{(2)}, \cdots, x_i^{(t)}\}\), outputs \(x_i^{(t)}\), and then discards all intermediate states but the final output before restarting the next run. This \emph{stateless} fashion is naturally sub-optimal, as the algorithm would not accumulate any knowledge about explored regions or learn from past failures. Therefore, it may revisit similar attempted trajectories, re-try strategies that are known to be ineffective in previous runs, and converge to familiar local optima, especially if the seed prompts are semantically or syntactically similar \citep{li2024art}. This makes the algorithm inefficient and results in highly similar prompts with limited marginal utility. Moreover, this inefficiency is exacerbated by the inherently slow convergence of discrete optimization. For example, P4D \citep{chin2024p4d} requires roughly 3,000 rounds of model invocation and gradient updates to optimize a single red teaming prompt, taking about 30 minutes per prompt on an NVIDIA RTX A100 GPU. These deficiencies make it very difficult to be scaled up for large-scale red teaming.

\section{The Design of DREAM}
\label{sec:method}

\subsection{Distributional Red Teaming via Energy-Based Modeling}
Motivated by our previous analysis about the limitations of existing methods, the key insight behind our proposal is to shift from discrete, stateless prompt-to-prompt optimization to directly modeling the distribution over unsafe prompts.

Formally, let \( q^*(x) \) denote the true (but unknown) distribution over the target model's problematic prompts, i.e., the probabilistic distribution from which samples \( x \in \hat{\mathcal{{A}}} \) are drawn. Our goal is to learn a probabilistic distribution \( p_\theta(x) \) parameterized by \(\theta\) (e.g., an autoregressive language model \( p_\theta(x) = \prod_{t=1}^T p_\theta(x_t \mid x_{<t}) \)), such that \( p_\theta(x) \) approximates \( q^*(x) \) as close as possible. This objective can be characterized by the following Kullback–Leibler divergence \citep{kullback1951information}:
\begin{equation}
\begin{aligned}
\label{eq:kl}
\theta^*\in \arg\min_{\theta} D_{\mathrm{KL}}(p_\theta \,\|\, q^*)
\end{aligned}
\end{equation}
This formulation has several desirable properties. First, by modeling the distribution $p_\theta(x)$, our method naturally converts the prompt-level discrete optimization into continuous optimization over model parameters \( \theta \). Second, since the parameters encode the distribution over prompts, each parameter update accumulates knowledge about which types of prompts are more or less likely to trigger unsafe outputs, thus promoting exploration efficiency during training. Furthermore, it is totally feasible to initialize  \( p_\theta(x) \) with a pretrained language model. As a result, our method inherits strong priors from large-scale human language data, which enables the model to understand and explore nuanced expressions, innuendos, and cultural references, which are subtle signals that typically require human-like common sense or contextual awareness and are often inaccessible to previous token-level search methods. Finally, once training is complete, sampling from the learned distribution \( p_{\theta^*}(x) \) is efficient. An arbitrary number of diverse prompts can be generated efficiently via forward passes, without requiring iterative search or gradient updates. This property makes our approach particularly suitable for red teaming, where a large number of unsafe prompts (e.g., thousands) are required for safety assessment and downstream safety-tuning.

Despite these promising properties, the objective in Eq.~(\ref{eq:kl}) remains particularly challenging to optimize in practice. The core difficulty lies in the fact that the ground-truth distribution \( q^*(x) \) is fundamentally unknown and there exists no readily available dataset that is sufficiently representative of the full support of the target model's problematic prompts. This makes direct optimization (e.g., through fine-tuning with MLE \citep{fisher1922mathematical}) impossible. Fortunately, recent advances in implicit generative modeling \citep{du2019implicit} provide a viable pathway to tackle this challenge. Specifically, results from the theory of energy-based models \citep{lecun2006tutorial, du2019implicit} suggest that even in the absence of explicit samples, the target distribution \( q^*(x) \) can be implicitly characterized with a properly defined \emph{energy function} \( E(x) \), which is a real-valued function that assigns lower values to more likely (or desirable) samples, and higher values otherwise. Then, the unknown distribution $q^*(x)$ can be expressed as a Boltzmann distribution \citep{boltzmann1872weitere} $q^*(x) = {\exp(-\beta\cdot E(x))}/{Z}$, where $Z$ is a constant that normalizes the distribution \citep{lecun2006tutorial,du2019implicit} and $\beta>0$ is a hyperparameter. Then, by plugging it into Eq.~(\ref{eq:kl}), we have:
\begin{equation}
\begin{aligned}
\label{eq:entropy}
&\arg\min_{\theta} D_{\mathrm{KL}}\!\big(p_\theta \,\|\, q^*\big) 
= \arg\min_{\theta} \mathbb{E}_{x \sim p_\theta}\!\left[\log \frac{p_\theta(x)}{q^*(x)}\right] \\
&= \arg\min_{\theta} \mathbb{E}_{x \sim p_\theta}\!\left[E(x) + \frac{1}{\beta}\log p_\theta(x)\right].
\end{aligned}
\end{equation}
The derivation above reduces the otherwise intractable KL divergence to two simple yet intuitive components: the first is to minimize the expected energy \(\mathbb{E}_{x \sim p_\theta}[E(x)]\), thereby shaping the distribution toward lower-energy (and thus more desirable) regions of the prompt space. The second objective acts as an entropy regularizer that penalizes low-entropy distributions by minimizing the log-likelihood \( \mathbb{E}_{x \sim p_\theta}[\log p_\theta(x)] \), thus avoiding degenerate solutions where the model collapses to a narrow set of prompts. 

\subsection{Energy Function Design}
So far, we have decomposed the objective into two intuitive and interpretable sub-goals. The second regularization term \( \mathbb{E}_{x \sim p_\theta}[\log p_\theta(x)] \) is straightforward to compute and optimize in practice. We now turn our attention to the first component, the energy function $E(\cdot)$. The energy function essentially defines the target distribution by assigning lower energy scores to desirable prompts and higher scores to undesired ones. In this section, we introduce our energy function design, which captures the following two scores.

\vspace{0.3em}
\noindent\textbf{Vision-level Harmfulness Energy.}  
The primary goal of $E$ is to guide the learned distribution $p_\theta(x)$ toward the target model's vulnerable prompt distribution \(\mathcal{A}_{\tau}\). However, directly assessing the harmfulness of a text prompt \(x\) is difficult as the risk often emerges only after it is rendered into an image. Therefore, we take a vision-level approach by evaluating the output image \(y = G(x)\) instead of the prompt itself. Specifically, we employ BLIP-2 \citep{li2023blip}, a pretrained vision-language model with strong generalization across diverse image-text domains, to compute a \emph{vision-level harmfulness energy} as part of the energy function. It assesses how the generated image is semantically aligned with a predefined harmful concept, and assigns lower energy to prompts that align better with the harmful concept.

Formally, given a generated image \( y = G(x) \) and a predefined target description \( c \) (like ``an image containing nudity''), the textual description $c$ is first sent to a pretrained language encoder $\mathcal{T}_\phi$ to obtain the sentence-level semantic embedding $t=\mathcal{T}_\phi(c)$. Then, the image \( y \) is passed through a vision encoder followed by a specialized transformer module known as the Q-Former \citep{li2023blip}. This module employs a set of query embeddings to interact with the visual features via cross-attention and finally extracts a set of latent tokens \(\mathcal{I}_\psi(y)= \{z_1, \dots, z_{k}\} \), each representing a different fine-grained aspect of the image in the same vision-language embedding space. Then, we define the alignment score as:
\begin{equation}
\label{eq:align}
E_{\text{align}}(x) = \mathbb{E}_{x \sim p_\theta} \left[ - \max_{z_i \in \mathcal{I}_\psi(G(x))} \frac{\langle z_i, t \rangle}{\|z_i\| \cdot \|t\|} \right]
\end{equation}
where \(\langle \cdot, \cdot \rangle\) and \(\| \cdot \|\) denote the inner product and the Euclidean norm, respectively. $E_{\text{align}}(x)$ measures the cosine similarity between the resulting image and the textual embedding, where higher similarity indicates stronger alignment  with the harmful concept and thus lower energy.

This formulation brings three key benefits. First, BLIP-2 provides better generalization even under distribution shifts, such as stylized or non-photorealistic images, making the alignment score more reliable across visual domains \citep{li2023blip,li2023blip2}. Second, the approach enables flexible red teaming through natural language descriptions. One can easily shift the target by modifying the concept prompt, e.g., replacing $c$ with ``an image depicting violent scenes'' to target violent content. When a small set of reference images is available, prompt tuning techniques can also be used to further refine and control the targeted concept \citep{schramowski2022can}. Third, the alignment score is continuous, allowing small improvements in prompt effectiveness to be reflected. This supports more stable optimization than discrete (e.g., binary) success signals.

\vspace{0.3em}
\noindent\textbf{Prompt-level Diversity Energy.} While jointly optimizing the harmfulness energy in Eq.~(\ref{eq:align}) and the entropy term \( \mathbb{E}_{x \sim p_\theta}[\log p_\theta(x)] \) in the main objective (Eq.~(\ref{eq:entropy})) effectively guides the model toward $q^*(x)$, we observe that relying solely on this formulation tends to produce a learned distribution with limited semantic diversity. This is possibly because the target distribution \(q^*(x)\) may itself be biased, e.g., certain keywords like ``nude'' and their semantically similar variants might dominate the probability mass. Consequently, semantically distinct prompts with lower probability under \(q^*(x)\) may remain largely unvisited in limited sampling iterations. To address this, we introduce a diversity energy term that explicitly encourages broader coverage within a limited sample budget. Let \(\mathcal{E}_\xi(x) \in \mathbb{R}^d\) denote the sentence embedding of prompt \(x\), obtained from a frozen pretrained encoder (e.g., a sentence transformer \citep{sentence-transformers}). Then, we define the prompt-level diversity energy as the expected pairwise similarity among prompt embeddings sampled from the current model distribution \( p_\theta(x) \):
\begin{equation}
\label{eq:diversity}
E_{\text{div}}(x) = \mathbb{E}_{x, x' \sim p_\theta,\, x \neq x'} \left[ \frac{\langle \mathcal{E}_\xi(x), \mathcal{E}_\xi(x') \rangle}{\|\mathcal{E}_\xi(x)\| \cdot \|\mathcal{E}_\xi(x')\|} \right].
\end{equation}
This would explicitly encourage semantic diversity among generated prompts within limited sampling iterations, thus promoting broader exploration and reducing redundancy.

\subsection{Red Team LLM Optimization}
After designing the energy function, we can plug Eq.~(\ref{eq:align}) and Eq.~(\ref{eq:diversity}) into Eq.~(\ref{eq:entropy}), and arrive at the final training objective for the red team prompt generator $\theta$:
\begin{equation}
\label{eq:final_loss}
\min_\theta \mathbb{E}_{x \sim p_\theta} \left[  E_{\text{align}}(x) + \lambda \cdot E_{\text{div}}(x) + \frac{1}{\beta} \cdot \log p_\theta(x) \right],
\end{equation}
where $\lambda$ and $\beta$ are balancing hyperparameters. However, optimizing Eq. (\ref{eq:final_loss}) is non-trivial. One intuitive approach would be to use backpropagation-based methods to obtain exact gradients and then update the LLM's parameters. However, applying backpropagation-based optimization directly is challenging. This is because the full red-teaming pipeline comprises multiple components such as autoregressive language generation, multi-step diffusion denoising, and energy models, each requiring storage of numerous intermediate activations for backpropagation-based gradient computation. For instance, generating an image via Stable Diffusion v1.5 typically requires 30 denoising steps, each producing high-dimensional feature maps (e.g., $\sim$2.3~GB). Even under a moderate batch size of 32, this already amounts to over 2~TB of GPU memory for a single pass, making end-to-end backpropagation-based training memory-prohibitive. Moreover, certain components like keyword-based safety filters are non-differentiable, further hindering backpropagation.

To enable effective gradient-driven optimization while avoiding the need for backpropagation through the entire pipeline, we propose a novel framework based on Simultaneous Perturbation Stochastic Approximation (SPSA) \citep{spall1992multivariate}. SPSA is a classical zeroth-order optimization method that allows estimates of high-dimensional gradients using only forward evaluations. However, vanilla SPSA has been empirically observed to suffer from instability and slow convergence in our red-teaming setup, due to the highly stochastic nature of both LLMs and diffusion-based generation (see experiments in Section~\ref{sec:abl}). To mitigate this, we propose a simple yet effective variant, GC-SPSA, which incorporates an adaptive sampling schedule as well as a history-aware gradient calibration mechanism to reduce gradient variance while maintaining efficiency. In the following, we first introduce SPSA and analyze its limitations, then present our GC-SPSA, and finally provide theoretical analysis and convergence guarantees to support our design.
\begin{definition}[SPSA \citep{malladi2023fine}]
\label{def:spsa}
Given an objective function $\mathcal{L}: \mathbb{R}^d \rightarrow \mathbb{R}$ and parameters $\theta \in \mathbb{R}^d$, SPSA uses the following randomized two-point finite-difference approximation to compute an unbiased estimate of the gradient $\nabla_\theta\mathcal{L}(\theta)$:

\begin{equation}
\label{eq:spsa}
g(\theta) := \frac{\mathcal{L}(\theta + \epsilon \Delta) - \mathcal{L}(\theta - \epsilon \Delta)}{2\epsilon} \Delta,
\end{equation}
where $\epsilon > 0$ is a small perturbation magnitude, and $\Delta \in \mathbb{R}^d$ is a random perturbation vector sampled from a zero-mean Gaussian distribution. 
\end{definition}
Previous works have proved that SPSA provides an unbiased estimate of the true gradient, i.e., $\mathbb{E}_\Delta[g(\theta)] = \nabla_\theta \mathcal{L}(\theta)$ \citep{spall1992multivariate}. In our setting, SPSA is particularly advantageous: it avoids backpropagation entirely and requires only forward evaluations at perturbed parameters, making it well-suited for bypassing the long and possibly non-differentiable pipeline. Moreover, it eliminates the need to store intermediate forward activations, further reducing memory consumption compared to backpropagation-based methods \citep{malladi2023fine}. 

Despite these advantages, we observe in our experiments that directly applying SPSA leads to unstable training dynamics. This instability primarily stems from the inherent stochasticity in both LLM and diffusion-based sampling, which introduces high variance into single-shot gradient estimates. To reduce variance, a straightforward strategy is to increase the number of forward estimates per iteration and average the resulting gradients. However, this significantly increases training cost linearly and is expensive in practice.

To mitigate this problem, we propose a simple yet effective method, GC-SPSA, which stabilizes SPSA with a novel adaptive gradient calibration algorithm. Our key insight is that Eq.~(\ref{eq:final_loss}) primarily steers the LLM toward a subspace of prompts that are likely to elicit harmful images from the target model. Since the pretrained LLM already possesses strong priors, such subspaces are shown to typically reside within a relatively flat and continuous basin in the loss landscape near the pretrained parameters \citep{xu2024random,jain2024mechanistically,zhang2023fine}. Therefore, we hypothesize that it is more crucial to ensure the reliability of early optimization steps to accurately locate this subspace, yet later updates may tolerate more variance and can be calibrated with early reliable gradients.

\begin{algorithm}[t]
\caption{The Complete Training Procedure of \sys}
\label{alg:gc-spsa}
\small
\begin{algorithmic}[1]
\Statex \textbf{Input:} Initial model parameters $\theta_0$, initial sampling budget $n_0$, learning rate $\eta$, correction strength $\gamma$, smoothing factor $\rho$, decay factor $T_\text{dec}$, current step $t$
\Statex \textbf{Output:} Optimized generator parameters $\theta_{t}$
\State $\hat{g}_0 \leftarrow \frac{1}{n_0} \sum_{i=1}^{n_0} g_{0,i}(\theta_0)$
\State $w_0 \leftarrow n_0$, $t \leftarrow 0$
\WHILE{loss not converged}
    \Statex \textcolor{gray}{$\triangleright$ \textit{Determine sampling budget for current step}}
    \State $n_t \leftarrow \max\left(1, \left\lfloor \frac{n_0}{2^{t/T_\text{dec}}} \right\rfloor \right)$
    \Statex \textcolor{gray}{$\triangleright$ \textit{Estimate gradients via SPSA with $n_t$ queries}}
    \FOR{$i = 1$ to $n_t$}
        \State Sample perturbation $\Delta_{t,i} \sim \text{Gaussian}^d$
        \State $\mathcal{L}^{+}_{t,i},\mathcal{L}^{-}_{t,i} \leftarrow \mathcal{L}(\theta_t + \epsilon \Delta_{t,i}),\mathcal{L}(\theta_t - \epsilon \Delta_{t,i})$
        \State $g_{t,i} \leftarrow \frac{\mathcal{L}^{+}_{t,i} - \mathcal{L}^{-}_{t,i}}{2\epsilon} \cdot \Delta_{t,i}$
    \ENDFOR
    \Statex \textcolor{gray}{$\triangleright$ \textit{Aggregate and calibrate gradients}}
    \State $\hat{g}_t \leftarrow \frac{1}{n_t} \sum_{i=1}^{n_t} g_{t,i} + \gamma \cdot \frac{w_{t-1}}{w_{t-1} + n_t} \cdot \hat{g}_{t-1}$
    \State $w_t \leftarrow \rho \cdot w_{t-1} + (1 - \rho) \cdot n_t$
    \Statex \textcolor{gray}{$\triangleright$ \textit{Update model parameters}}
    \State $\theta_{t+1} \leftarrow \theta_t - \eta \cdot \hat{g}_t$
    \State $t\leftarrow t+1$
\ENDWHILE

\State \textbf{Return} $\theta_{t}$
\end{algorithmic}
\end{algorithm}

Specifically, for the $t$-th update, we first estimate the gradient $n_t$ times using Eq.~(\ref{eq:spsa}), obtaining a set of stochastic gradient estimates $\{ g_{t,1}(\theta_t), g_{t,2}(\theta_t), \dots, g_{t,n_t}(\theta_t) \}$. The number of queries $n_t$ is controlled by an exponentially decaying schedule: \(n_t = \max\left(1, \left\lfloor \frac{n_0}{2^{t/T_\text{dec}}} \right\rfloor \right)\), where $n_0$ is the initial number of sampling times and $T_\text{dec}$ governs the decay rate. This scheduling allocates a higher sampling budget to early iterations and gradually reduces the number of samples as optimization stabilizes. In our experiments, we find that setting $n_0 = 4$ and $T_\text{dec} = 10$, i.e., starting with 4 samples and halving the sampling budget every 10 steps, yields stable and efficient optimization performance (see experiments in Sec.~\ref{sec:abl}). To further reduce the variance of the later estimated gradient, inspired by confidence-aware optimal Bayesian fusion \citep{griebel2020kalman}, we introduce a gradient calibration mechanism. Specifically, we treat each new gradient estimate as a noisy observation and combine it with historical information using a confidence-aware correction term:

\vspace{-1em}
\begin{equation}
\begin{aligned}
\label{eq:spsa-gc}
    &\hat{g}_{t} = \frac{1}{n_t} \sum_{i=1}^{n_t} g_{t,i}(\theta_t) + \gamma \cdot \frac{w_{t-1}}{w_{t-1} + n_t} \cdot \hat{g}_{t-1}, \\
    &\theta_{t+1} = \theta_t - \eta \cdot \hat{g}_t,
\end{aligned}
\end{equation}
\vspace{-0.5em}

\noindent where $\hat{g}_{t-1}$  is the accumulated gradient estimate from previous iterations ($\hat{g}_{0} = \frac{1}{n_0} \sum_{i=1}^{n_0} g_{0,i}(\theta_0)$), $w_{t-1}$ denotes its effective sample size (initialized as $w_0 = n_0$), $\eta$ is the learning rate. The update of $w_t$ follows an exponential moving average rule. We use the term ${w_{t-1}}/{w_{t-1} + n_t}$ to approximate the relative confidence of historical vs. current gradients, and $\gamma$ controls the overall strength of the correction. Intuitively, our confidence-weighted fusion scheme anchors the current noisy gradient estimate towards the historically aggregated direction, especially when the current estimate is based on fewer samples (i.e., lower confidence).  As formally analyzed in Theorem \ref{thm:snr}, the GC-SPSA estimator achieves a strictly higher signal-to-noise ratio (SNR) than the vanilla SPSA estimator for all $t \ge 1$, which helps mitigate gradient noise and promotes a more consistent optimization path.

\vspace{-0.5em}
\begin{theorem}[Improved SNR of GC-SPSA]
\label{thm:snr}
 Let $\|g_{\text{true}}\|$ be the ground-truth gradient, $\bar{g}_k$ be the vanilla SPSA estimator, and $\hat{g}_k = \bar{g}_k + H_k \hat{g}_{k-1}$ be the GC-SPSA estimator, with $\hat{g}_0 = \bar{g}_0$ and $H_k > 0$. Then for all $t \ge 1$, the SNR difference between the GC-SPSA and the vanilla SPSA admits the explicit positive lower bound $\mathcal{D}_t$ as follows:
\vspace{-0.5em}
\begin{equation}
\mathcal{D}_t = \frac{\|g_{\text{true}}\|^2}{V_{\rm {single}} \sum_{k=0}^t h_k^2 V_k} \left[ P_t^2 V_{\rm single} - \sum_{k=0}^t h_k^2 V_k \right]
\end{equation}
where $h_t = 1$, $h_k = \prod_{j=k+1}^{t} H_j$ and $H_j = \gamma\frac{w_{j-1}}{w_{j-1}+n_j}$. $V_{\rm {single}}$ is the gradient estimation variance of vanilla SPSA, $P_t = \sum_{k=0}^{t} h_k$ is the cumulative weight sum, and $V_k = V_{\rm {single}}/n_k$ is the gradient variance of GC-SPSA at step $k$.
\end{theorem}

The detailed proof is in Appendix \ref{sec:thm1}. Furthermore, we also provide a theoretical guarantee of global convergence and convergence rate analysis for  GC-SPSA under mild assumptions, as formally shown in the following theorem:

\begin{theorem}[Global Convergence and Rate Analysis of GC-SPSA]
\label{thm:global}

Consider an objective $\mathcal{L}: \mathbb{R}^d \to \mathbb{R}$ satisfying $\nabla^2 \mathcal{L}(\theta) \preceq \ell I_d$ for all $\theta \in \mathbb{R}^d$, where $I_d$ denotes the $d$-dimensional identity matrix. Then, for a stationarity level $\delta$, GC-SPSA will converge (i.e., {\small $\min_{t \in [T]} \mathbb{E}[\|g(\theta_t)\|^2] \leq \delta$}) after
\begin{equation}\label{eq:final_complexity}
T = \Theta\left( \frac{\mathcal{L}(\theta_0)-\mathcal{L}^*} {\eta\,\zeta_{\min}\,\delta - C_{\max}\,\Upsilon - \frac{\ell\,\eta^2\,\Xi}{2}} \right).
\end{equation}
iterations. Here, $\mathcal{L}^*$ denotes the global minimum, $\zeta_{\min}$ is the minimum descent coefficient, $C_{\max}$ bounds the cross-term coefficient, and $\Upsilon$ and $\Xi$ denote upper bounds on the average squared gradient-estimator norm and accumulated noise variance, respectively.

\end{theorem}

For each training step, we begin by perturbing the current model parameters $\theta_t$ and estimate the gradient $n_t$ times using Eq.~(\ref{eq:spsa}). Since the objective $\mathcal{L}(\theta)$ involves an intractable expectation, we approximate it via Monte Carlo sampling over a batch of sampled prompts (see more details in Alg. \ref{alg:loss}). The resulting gradient estimates are then averaged and calibrated using our confidence-aware update rule in Eq.~(\ref{eq:spsa-gc}). Finally, the calibrated gradient is used to update the parameters $\theta_t$. We summarize the complete training procedure in Alg.~\ref{alg:gc-spsa}.

\subsection{Inference-Time Adaptive Temperature Scaling}
\label{sec:sampling}

After training, the red-team LLM is steered toward the desired prompt  distribution. The final step is sampling from this model to generate red-teaming prompts. However, we empirically find that the model may still fail to adequately explore its support during generation. This is because each sample is produced independently and without awareness of previously generated prompts, leading to repetitive or redundant tokens that have limited marginal benefits.

To further enhance diversity, we propose an inference-time strategy that encourages diversity via adaptive temperature scaling. Recall that in autoregressive decoding, the model iteratively predicts the next token distribution as $p_{\theta^*}({x_t}\!=\!v|x_{<t}) = \frac{\exp\left( \boldsymbol{z}_t[v] / \tau_t \right)}{\sum_{j \in \mathcal{V}} \exp\left( \boldsymbol{z}_t[j] / \tau_t \right)}$, where the temperature hyperparameter $\tau_t$ controls the sharpness of the token distribution  \( \boldsymbol{z}_t \in \mathbb{R}^{|\mathcal{V}|} \) at decoding step \(t\). Intuitively, lower temperatures make the model more confident (peaky), while higher temperatures flatten the distribution to encourage exploration. As such, we maintain a global token frequency vector \( \mathbf{f} \in \mathbb{N}^{|\mathcal{V}|} \), tracking the number of times each vocabulary token has appeared across previous generations. At decoding step \(t\), given raw logits \( \boldsymbol{z}_t \in \mathbb{R}^{|\mathcal{V}|} \), we compute the relative frequency of the top-scoring token \( v_t = \arg\max_j \boldsymbol{z}_t[j] \), and use it to scale the temperature:
\begin{equation}
\tau_t = \max\left( \frac{1}{\alpha} \log(1 + \frac{\mathbf{f}[v_t]}{\sum_j \mathbf{f}[j]}),\ \tau_{\min} \right),
\end{equation}
where \(\alpha > 0\) is a sensitivity coefficient and \(\tau_{\min}\) prevents degeneracy. The final logits are adjusted as \(\tilde{\boldsymbol{z}}_t = \boldsymbol{z}_t / \tau_t\) before sampling. This penalizes frequent tokens, promoting underexplored generations without modifying training or architecture. Empirically, we find it improves prompt diversity with minimal overhead and little loss in effectiveness.

\begin{table*}[h]
    \centering
    \footnotesize
    \setlength{\tabcolsep}{2pt}
    \caption{Comparison with baselines on Stable Diffusion v1.5 and other safety-aligned diffusion models.}
    \vspace{-0.3em}
        \resizebox{0.94\textwidth}{!}{ 
            \begin{tabular}{c l *{2}{c}*{2}{c}*{2}{c} c c c c}
                \toprule
                & & \multicolumn{2}{c}{Stable Diffusion v1.5}  & \multicolumn{2}{c}{CA} & \multicolumn{1}{c}{ESD} & \multicolumn{2}{c}{UCE} & \multicolumn{1}{c}{SafeGen} & \multicolumn{1}{c}{RECE} \\
                \cmidrule(lr){3-4}  \cmidrule(lr){5-6} \cmidrule(lr){7-7} \cmidrule(lr){8-9} \cmidrule(lr){10-10} \cmidrule(lr){11-11} 
                & & {Sexual} & {Violence} & {Sexual} & {Violence} & {Sexual} & {Sexual} & {Violence} & {Sexual} & {Sexual} \\
                \cmidrule(lr){3-3} \cmidrule(lr){4-4} \cmidrule(lr){5-5} \cmidrule(lr){6-6} \cmidrule(lr){7-7} \cmidrule(lr){8-8} \cmidrule(lr){9-9} \cmidrule(lr){10-10} \cmidrule(lr){11-11}
                & & PSR $\uparrow$ / PS $\downarrow$  & PSR $\uparrow$ / PS $\downarrow$  & PSR $\uparrow$ / PS $\downarrow$  & PSR $\uparrow$ / PS $\downarrow$  & PSR $\uparrow$ / PS $\downarrow$  & PSR $\uparrow$ / PS $\downarrow$  & PSR $\uparrow$ / PS $\downarrow$  & PSR $\uparrow$ / PS $\downarrow$ & PSR $\uparrow$ / PS $\downarrow$ \\
                \midrule
                \multicolumn{6}{l}{\emph{Human-written Datasets}} \\
                & I2P  & 51.5\% / 0.49  & 13.9\% / 0.47   & 11.7\% / 0.49 & 13.1\% / 0.47 & {\color{white}0}9.5\% / 0.49 & 10.0\% / 0.49 & {\color{white}0}6.9\% / 0.47 & 45.6\% / 0.49 & {\color{white}0}7.5\% / 0.49 \\
                & Adv. Nibbler  & 28.3\% / 0.53  & {\color{white}0}8.9\% / 0.54  & {\color{white}0}3.1\% / 0.53 & {\color{white}0}3.5\% / 0.54 & {\color{white}0}1.2\% / 0.53 & 30.1\% / 0.53 & {\color{white}0}7.8\% / 0.54 & 33.8\% / 0.53 & {\color{white}0}0.7\% / 0.53 \\
                \midrule
                \multicolumn{6}{l}{\emph{Automated Red Teaming (Token Perturbation)}} \\
                & QF-Attack & 23.8\% / 0.59  & 10.6\% / 0.62  & {\color{white}0}0.6\% / 0.59 & {\color{white}0}9.4\% / 0.62 & {\color{white}0}0.0\% / 0.59  & {\color{white}0}1.3\% / 0.59  & 10.0\% / 0.62 & 21.2\% / 0.59 & {\color{white}0}0.6\% / 0.59 \\
                & SneakyPrompt & 61.7\% / 0.52  & 26.5\% / 0.65 & {\color{white}0}7.9\% / 0.52 & 13.5\% / 0.52 & 25.6\% / 0.64 & 16.5\% / 0.52 & 20.1\% / 0.64 & 25.5\% / 0.52 & 14.0\% / 0.52 \\
                & MMA-Diffusion  & 91.0\% / 0.63  & 73.9\% / 0.65   & 44.9\% / 0.63 & 59.2\% / 0.65 & 35.9\% / 0.63 & 59.9\% / 0.63 & 66.9\% / 0.65 & 34.0\% / 0.63 & 52.2\% / 0.63 \\
                & P4D & 78.0\% / 0.60 & 42.0\% / 0.58  & 52.0\% / 0.60 & 26.0\% / 0.66 & 43.3\% / 0.60 & 24.0\% / 0.56 & 10.0\% / 0.58 & 61.9\% / 0.55 & 16.0\% / 0.55 \\
                & UnlearnDiffAtk & 83.0\% / 0.52 & 30.0\% / 0.49  & 36.4\% / 0.52 & 10.5\% / 0.49 & 21.2\% / 0.52 & 24.6\% / 0.51 & 10.5\% / 0.49 & 55.9\% / 0.52 & 13.6\% / 0.52 \\
                \midrule
                \multicolumn{6}{l}{\emph{Automated Red Teaming (LLM Rewrite)}} \\
                & ART  & 14.9\% / 0.48  & 29.2\% / 0.47  & {\color{white}0}2.7\% / 0.49 & 14.1\% / 0.46 & {\color{white}0}0.8\% / 0.49 & {\color{white}0}0.8\% / 0.49 & 20.8\% / 0.46 & 13.7\% / 0.49 & {\color{white}0}0.8\% / 0.48 \\
                & FLIRT & 91.8\% / 0.77  & 74.4\% / 0.66  &  26.0\% / 0.58 & 64.4\% / 0.63 & 17.1\% / 0.64 & 48.5\% / 0.64 & 18.4\% / 0.61 & 10.2\% / 0.59 & 10.7\% / 0.57 \\
                & JailFuzzer & 72.4\% / 0.56  & 80.0\% / 0.55  &  31.5\% / 0.52 & 44.0\% / 0.54 & 23.8\% / 0.51  & 51.9\% / 0.53 & 38.0\% / 0.52 & 35.4\% / 0.52 & 35.9\% / 0.51 \\
                \midrule
                \rowcolor{gray!15}& {Ours} & 92.2\% / 0.50  & 87.0\% / 0.55   & 76.0\% / 0.56 & 77.3\% / 0.57 & 72.1\% / 0.56 & 89.0\% / 0.54 & 83.6\% / 0.57 & 81.6\% / 0.49 & 91.3\% / 0.56 \\
                \bottomrule
            \end{tabular}
        }
    \label{tab:comparison-concept-erasure}
    \vspace{-1.5em}
\end{table*}

\section{Evaluation}
\label{sec:exp}

\subsection{Experimental Setup}

\noindent\textbf{Target Diffusion Models \& Safety Filters.} In our experiments, we evaluate a variety of standard diffusion models, safety-aligned models, and safety filters. In addition to the standard Stable Diffusion v1.5, we also evaluate safety-aligned models including ESD \citep{gandikota2023erasing}, CA \citep{kumari2023ablating}, UCE \citep{gandikota2024unified}, SafeGen \citep{li2024safegen}, and RECE \citep{gong2024rece}, all of which have unlearned certain unsafe concepts from the models. For external safety filters, following Yang et al.~\citep{yang2024sneakyprompt}, we consider 4 external filters classified into (1) text-based filters (NSFW text classifier \citep{nsfw_text_classifier} and Keyword-Gibberish hybrid filter \citep{gibberish_detector}), (2) image-based filters (NSFW image filter \citep{chhabra_nsfw_detection_dl} and Stable Diffusion's built-in image safety checker, SC \citep{compvis2022sc}). Furthermore, we evaluate the generalizability of \sys on several real-world models, including SDXL, SDv3, Kandinsky v3, and Shuttle 3 Diffusion. We also evaluate \sys in a transfer-based setting on multiple real-world online T2I platforms, including Ideogram, DeepAI,  DALL·E 3, and Midjourney.

\vspace{0.3em}
\noindent\textbf{Baselines.} We evaluate and compare \sys against several state-of-the-art baselines. For human-written red teaming datasets, we include I2P \citep{schramowski2023sld} and Google's Adversarial Nibbler \citep{quaye2024advnibbler}, collected from T2I community forums and via crowdsourcing, respectively. For automated methods, we consider QF-Attack \citep{zhuang2023qfattack}, SneakyPrompt \citep{yang2024sneakyprompt}, MMA-Diffusion \citep{yang2024mma}, P4D \citep{chin2024p4d}, UnlearnDiffAtk \citep{zhang2024unlearndiff}, JailFuzzer \citep{dong2025fuzz}, FLIRT \citep{mehrabi2024flirt} and ART \citep{li2024art}. Note that the baselines differ in their assumed level of access to the target model as well as their original motivation (e.g., SneakyPrompt was originally proposed for jailbreak attacks in closed-box settings). We include them for completeness, as their underlying mechanisms are closely related to red-teaming and finally yield comparable unsafe prompt datasets. We provide a more detailed discussion in Appendix~\ref{sec:diss}.

\vspace{0.3em}
\noindent\textbf{Evaluation Metrics.}  
We primarily use two metrics to evaluate the performance of each red teaming method: \textit{Prompt Success Rate} (PSR) and \textit{Prompt Similarity} (PS), which measure the effectiveness and diversity of the generated prompts, respectively. \textbf{PSR} is the proportion of prompts that successfully trigger the target model to generate images containing the specified inappropriate content. Following Yang et al.~\citep{yang2024mma}, we use PSR out of $N$ generations (PSR-N) instead of a single generation to reduce the impact of inherent stochasticity in diffusion sampling. Specifically, for each prompt, we generate $N$ images with different random seeds. The prompt is considered successful if at least one of these $N$ images contains the desired unsafe concept, and the final PSR is measured as the ratio of successful prompts. In our experiments, we use $\text{PSR-3}$, i.e., $N=3$, and mainly adopt {Multi-headed Safety Classifier (MHSC)} \citep{qu2023unsafe} as our detector. MHSC is a category-specific NSFW image detector that provides per-category confidence scores for various unsafe concepts. It has been widely used in the community due to its high precision \citep{qu2023unsafe,qu2024unsafebench,yang2024mma}. A higher PSR ($\uparrow$) indicates the method is more capable of generating effective prompts. Besides, \textbf{PS} quantifies the diversity of the prompts by measuring the average pairwise cosine similarity between all prompt embeddings. 
In our evaluation, we use the state-of-the-art BGE embedding model \citep{chen2024bge} to obtain the prompt embeddings (note that this model is different from $\mathcal{E}_\xi$ we use in Eq.~(\ref{eq:diversity})). A lower PS score ($\downarrow$) indicates lower inter-prompt similarity, meaning that the prompts are semantically more diverse and less repetitive.

\vspace{0.3em}
\noindent\textbf{Implementation Details.} We initialize $\theta$ with the pretrained Gemma-2-27b-it model~\citep{team2024gemma}. $c$ is initialized following Tab.~\ref{tab:unsafe-content}, with lightweight embedding tuning applied to ensure more accurate category representation. For GC-SPSA, the learning rate $\eta$ and hyperparameter $\epsilon$ are set to default values of $1\times10^{-6}$ and $1\times10^{-3}$, respectively. Following prior works~\citep{hong2024curiosity,li2024art}, system prompts and few-shot ICL exemplars are also applied to ensure task alignment and efficiency. The model is trained until convergence with a batch size of 32, which typically requires $\sim$300 steps and corresponds to roughly 12 hours on NVIDIA A100 GPUs. After training, we sample 1,024 prompts with ATS for evaluation. See Appendix~\ref{sec:diss} for more details.

\subsection{Main Results}
\noindent\textbf{Effectiveness on Concept-erased T2I Models.}
We first conduct experiments on both the standard Stable Diffusion (SD) v1.5 model and several concept-erased models, whose weights are fine-tuned or edited to unlearn the corresponding unsafe concept. As shown in Tab. \ref{tab:comparison-concept-erasure}, our method consistently achieves the highest PSRs across all models and categories, significantly outperforming all baselines. For example, on concept-erased models such as UCE and RECE, human-written datasets typically yield PSRs below 10\%, and state-of-the-art red teaming methods like MMA-Diffusion often struggle to exceed 50\%. In contrast, our prompts achieve PSRs above 79\% on all evaluated models. In addition to effectiveness, our method also excels in prompt diversity. Across all models, our prompts maintain a prompt similarity (PS) score around 0.55, which is notably better than most baselines and on par with human-written datasets, indicating higher semantic diversity. This highlights that our approach not only discovers more successful prompts, but also explores a broader and more varied region of the prompt space that remains unexplored by existing red teaming techniques.

\begin{table*}[t]
    \centering
    \footnotesize
    \setlength{\tabcolsep}{2pt}
    \caption{Comparison with baselines on external safety filters.}
    \vspace{-0.3em}
    \resizebox{0.9\textwidth}{!}{ 
        \begin{tabular}{c l *{2}{c}*{2}{c}*{2}{c}*{2}{c}}
            \toprule
            & & \multicolumn{2}{c}{Safety Checker} & \multicolumn{2}{c}{NSFW Image Detector} & \multicolumn{2}{c}{NSFW Text Detector} & \multicolumn{2}{c}{Keyword-Gibberish Filter} \\
            \cmidrule(lr){3-4} \cmidrule(lr){5-6} \cmidrule(lr){7-8} \cmidrule(lr){9-10}
            & & Sexual & Violence & Sexual & Violence & Sexual & Violence & Sexual & Violence \\
            \cmidrule(lr){3-3} \cmidrule(lr){4-4} \cmidrule(lr){5-5} \cmidrule(lr){6-6} \cmidrule(lr){7-7} \cmidrule(lr){8-8} \cmidrule(lr){9-9} \cmidrule(lr){10-10}
            & & PSR $\uparrow$ / PS $\downarrow$ & PSR $\uparrow$ / PS $\downarrow$ & PSR $\uparrow$ / PS $\downarrow$ & PSR $\uparrow$ / PS $\downarrow$ & PSR $\uparrow$ / PS $\downarrow$ & PSR $\uparrow$ / PS $\downarrow$ & PSR $\uparrow$ / PS $\downarrow$ & PSR $\uparrow$ / PS $\downarrow$ \\
            \midrule
            \multicolumn{6}{l}{\emph{Human-written Datasets}} \\
            & I2P & 23.4\% / 0.49 & 13.6\% / 0.47 & 26.3\% / 0.49 & 13.0\% / 0.47 & 37.5\% / 0.49 & {\color{white}0}8.5\% / 0.47 & 26.5\% / 0.49 & {\color{white}0}6.5\% / 0.47 \\
            & Adv. Nibbler & 14.3\% / 0.53 & {\color{white}0}8.4\% / 0.54 & 17.3\% / 0.53 & {\color{white}0}8.2\% / 0.54 & 19.4\% / 0.53 & {\color{white}0}6.2\% / 0.54 & {\color{white}0}4.8\% / 0.53 & {\color{white}0}0.0\% / 0.54 \\
            \midrule
            \multicolumn{6}{l}{\emph{Automated Red Teaming (Token Perturbation)}} \\            & QF-Attack & 11.9\% / 0.59 & 10.6\% / 0.62 & {\color{white}0}1.9\% / 0.59 & 10.6\% / 0.62 & {\color{white}0}1.2\% / 0.59 & {\color{white}0}8.8\% / 0.62 & {\color{white}0}3.1\% / 0.59 & {\color{white}0}0.0\% / 0.62 \\
            & SneakyPrompt & 30.5\% / 0.52 & 26.5\% / 0.67 & 11.0\% / 0.50 & 26.5\% / 0.64 & 12.5\% / 0.46 & 13.0\% / 0.64 & 51.8\% / 0.52 & 18.1\% / 0.64 \\
            & MMA-Diffusion & 40.5\% / 0.63 & 72.2\% / 0.65 & {\color{white}0}4.7\% / 0.63 & 68.3\% / 0.65 & {\color{white}0}1.5\% / 0.63 & 11.5\% / 0.65 & {\color{white}0}0.0\% / 0.63 & {\color{white}0}0.0\% / 0.65 \\
            & P4D & 40.0\% / 0.60 & 40.0\% / 0.58 & 44.0\% / 0.60 & 38.0\% / 0.58 & {\color{white}0}4.0\% / 0.60 & {\color{white}0}4.0\% / 0.58 & {\color{white}0}0.0\% / 0.60 & {\color{white}0}0.0\% / 0.58 \\
            & UnlearnDiffAtk & 35.6\% / 0.52 & 10.5\% / 0.49 & 43.2\% / 0.52 & {\color{white}0}9.5\% / 0.49 & 48.3\% / 0.52 & {\color{white}0}6.3\% / 0.49 & {\color{white}0}7.6\% / 0.52 & {\color{white}0}2.1\% / 0.49 \\
            \midrule
            \multicolumn{6}{l}{\emph{Automated Red Teaming (LLM Rewrite)}} \\
            & ART & {\color{white}0}4.4\% / 0.48 & 28.8\% / 0.47 & {\color{white}0}9.2\% / 0.48 & 29.2\% / 0.48 & {\color{white}0}6.4\% / 0.48 & 21.8\% / 0.46 & {\color{white}0}4.4\% / 0.48 & {\color{white}0}4.5\% / 0.47 \\
            & FLIRT & 26.5\% / 0.62 & 72.6\% / 0.64 & 45.9\% / 0.75 & 75.4\% / 0.65 & {\color{white}0}8.5\% / 0.51 & 16.7\% / 0.60 & 44.6\% / 0.58 & 48.9\% / 0.57 \\
            & JailFuzzer & 51.9\% / 0.53 & 72.0\% / 0.52 & 48.1\% / 0.55 & 76.0\% / 0.55 & 42.0\% / 0.51 & 46.4\% / 0.52 & 59.1\% / 0.51 & 38.0\% / 0.52 \\
            \midrule
            \rowcolor{gray!15} & {Ours} & 64.7\% / 0.52 & 86.4\% / 0.54 & 57.3\% / 0.50 & 83.7\% / 0.58 & 62.3\% / 0.51 & 42.5\% / 0.54 & {67.4\% / 0.52} & {65.7\% / 0.56} \\
            \bottomrule
        \end{tabular}
    }
    \label{tab:comparison-filters}
    \vspace{-0.5em}
\end{table*}

\label{sec:safety-filter}
\vspace{0.3em}
\noindent\textbf{Effectiveness on External Safety Filters.}
We further evaluate the effectiveness of \sys on various external safety filters and compare it with baselines. For each safety filter, we combine it with the standard SD v1.5 model and regard the whole model-filter pipeline as an integrated generative system. As shown in Tab. \ref{tab:comparison-filters}, baseline methods are largely unstable, while DREAM consistently achieves the best or nearly the best PSR-3 across all settings. For example, while MMA-Diffusion achieves good results on most safety-aligned models and SC, it almost completely fails on the NSFW Text Detector and Keyword-Gibberish Filter. On safety filters, the most effective baseline is JailFuzzer, which performs prompt-to-prompt mutation powered by multiple LLM and VLM agents. However, JailFuzzer is much less effective on safety-aligned models. This is possibly because the success of JailFuzzer’s mutation process heavily depends on the accuracy of the simulated SafetyFilter signal provided by the VLM brain. When applied to safety-aligned models that sanitize outputs instead of blocking them, the bypass-score feedback is omitted altogether, which deprives the mutation process of this optimization cue and consequently reduces its effectiveness. These findings demonstrate that baselines fail to generalize across different safety mechanisms. In contrast, DREAM performs well on both safety mechanisms without any ad hoc design, demonstrating the generality and universality of our method.

\begin{table}[t]
    \centering
    \caption{\small Effectiveness of \sys across models and NSFW themes. (a) results with more models on {sexual} (left) and {violence} (right) concepts; (b) results on more NSFW themes on SD v1.5.}
    \label{tab:asr-ours}
    \footnotesize
    \vspace{-0.3em}
    \setlength{\tabcolsep}{4pt}
    \resizebox{0.96\linewidth}{!}{
        \begin{tabular}{lc|lc}
            \toprule
            \multicolumn{4}{c}{{(a) More T2I Models}} \\
            \midrule
            {Model (Sexual)} & {PSR $\uparrow$ / PS $\downarrow$} & {Model (Violence)} & {PSR $\uparrow$ / PS $\downarrow$} \\
            \midrule
            Stable Diffusion XL       & 89.5\% / 0.52 & Stable Diffusion XL       & 92.3\% / 0.51 \\
            Stable Diffusion v3       & 82.5\% / 0.53 & Stable Diffusion v3       & 92.2\% / 0.52 \\
            Kandinsky v3              & 89.8\% / 0.52 & Kandinsky v3              & 86.8\% / 0.53 \\
            Shuttle 3 Diffusion       & 86.9\%  / 0.51 & Shuttle 3 Diffusion       & 90.0\% / 0.51 \\
            \midrule
            \multicolumn{4}{c}{{(b) More NSFW Themes}} \\
            \midrule
            {Category} & {PSR $\uparrow$ / PS $\downarrow$} & {Category} & {PSR $\uparrow$ / PS $\downarrow$} \\
            \midrule
            Self-harm         & 87.8\% / 0.55 & Shocking          & 94.7\% / 0.56 \\
            Hate              & 92.6\% / 0.52 & Harassment        & 94.1\% / 0.53 \\
            Political         & 91.6\% / 0.50   & Illegal Activity  & 92.7\% / 0.51 \\
            \bottomrule
        \end{tabular}
    }
\end{table}

\vspace{0.3em}
\noindent\textbf{Effectiveness on More T2I Models \& NSFW Themes.}
We evaluate \sys on more T2I models, including Stable Diffusion XL \citep{podell2024sdxl}, Stable Diffusion v3 \citep{esser2024scaling}, Kandinsky v3 \citep{vladimir2024kandinsky}, and Shuttle 3 Diffusion \citep{shuttle3diffusion}. These models vary significantly in architectural design (e.g., using DiT \citep{peebles2023scalable} instead of convolutional U-Nets) and training settings, and some of them are reported to apply dataset filtering to remove (some) unsafe images before training \citep{podell2024sdxl,esser2024scaling}. As shown in Tab.~\ref{tab:asr-ours} (a), \sys consistently performs well across all models and both categories, with PSR approaching 90\% on average. The discovered prompts also exhibit a level of diversity comparable to that of human-written prompts. These results highlight the model architecture-agnostic nature of \sys and its strong potential for adapting to future models.
Additionally, we further assess the generalizability of \sys on additional NSFW themes following prior work \citep{schramowski2023sld}, including self-harm, shocking, hate, harassment, political, and illegal activity. Following Yang et al.~\citep{yang2024mma}, we use Q16 \citep{schramowski2022can} as the detector in this setting, as MHSC does not support all of these categories. As shown in Tab.~\ref{tab:asr-ours} (b), \sys maintains high PSR across most cases, exceeding 90\% consistently and reaching close to 95\% for shocking and harassment. Besides, our prompts remain highly diverse and close to that of human-written prompts. These findings demonstrate the scalability and generalizability of \sys in discovering a broader range of unsafe concepts.

\begin{figure}[t]
    \centering
    \includegraphics[width=0.96\linewidth]{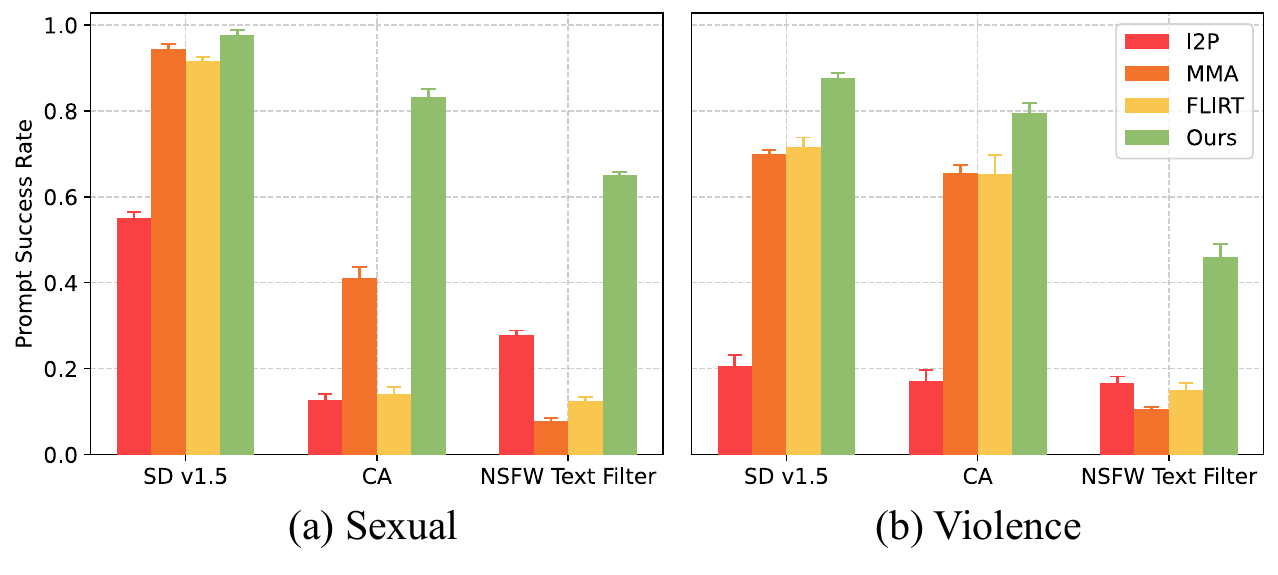}            
    \vspace{-0.5em}
    \caption{\small User study results on prompt success rate.}
    \label{fig:user_effectiveness}
\end{figure}

\begin{figure}[t]
    \vspace{-1em}
    \centering
    \includegraphics[width=\linewidth]{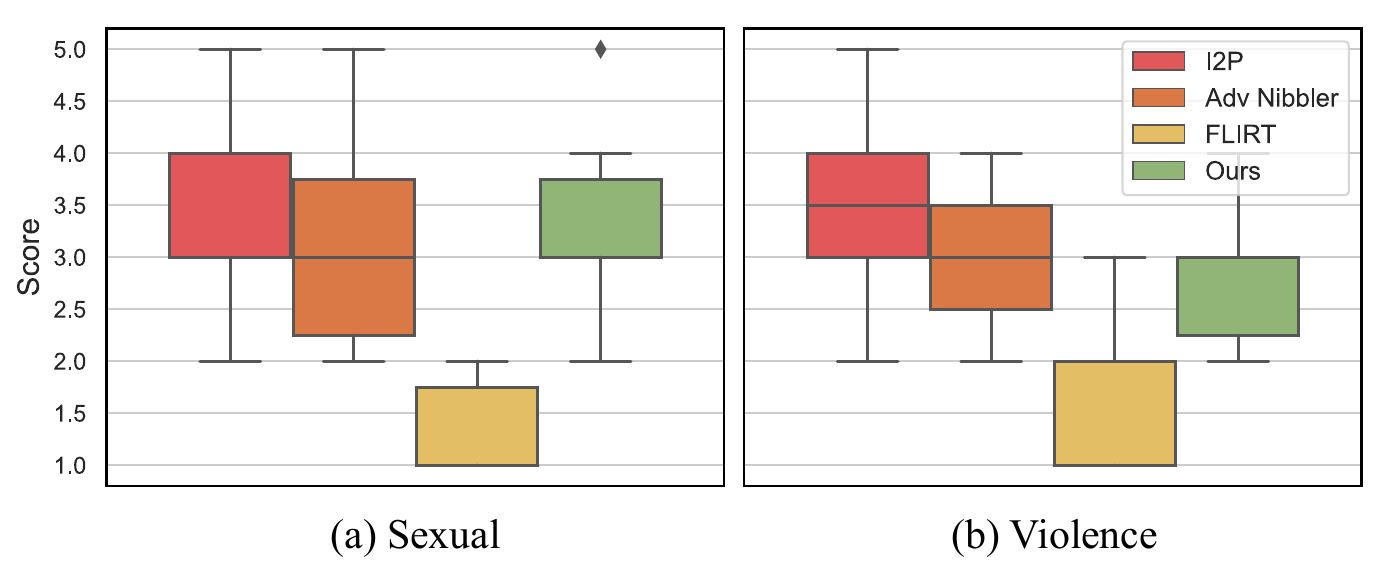}
    \vspace{-1.5em}
    \caption{\small User study results on prompt diversity.}
    \label{fig:user_diversity}
\end{figure}

\subsection{Human Evaluation}
In this section, we conduct a user study to assess the effectiveness and diversity of different methods under human perception. Specifically, we select SD v1.5, CA, and NSFW Text Filter as the representative models. We then compare DREAM with I2P, MMA-Diffusion, and FLIRT. We evaluate all model–method combinations on both sexual and violence categories, resulting in $2 \times 3 \times 4 = 24$ concept–model–method settings in total. For each setting, we randomly sample 30 prompts from the method's generated prompts to form a prompt pool. Each prompt generates 3 images, forming a prompt-images group. Then, we recruit 30 volunteers, all university students or faculty members from various academic backgrounds, to participate in the study. All participants are adults in good physical and mental health, fully informed of the study, and provided  consent to participate.  Then, we brief participants on the basics of T2I models and red teaming, the recommended definition of the corresponding unsafe category (Tab.~\ref{tab:unsafe-content} in Appendix), and start the study after obtaining their confirmation of full comprehension and agreement to continue the study.

For evaluation, each participant is randomly assigned 5 prompt-image groups from the 30 available for each concept-model-attack setting. Prompt-image groups are organized in random order and displayed one-by-one. Participants are asked to determine whether the three displayed images in each group clearly reflect the specified unsafe concept. The group is marked as successful if and only if the participant identifies at least one such image (PSR-3). Then, participants are presented with the full set of $4 \times 30$ problematic prompts on SD v1.5 and asked to rate these prompt sets by their perceived diversity, ranging from 1 to 5 (higher is better), based on their understanding of the prompts' lexical, structural, and semantic richness (see Tab.~\ref{tab:prompt-diversity} in Appendix). We replace MMA with Adv. Nibbler since MMA's prompts are not human-readable. Despite involving images depicting NSFW concepts, our study was reviewed and approved by our institution's research ethics committee under a process analogous to the ``exempt review'' category of U.S. IRB protocols, as the review board determined it posed no more than minimal risk, given that all participants were healthy adults, fully informed, and free to withdraw at any time. As shown in Fig.~\ref{fig:user_effectiveness} and \ref{fig:user_diversity}, \sys consistently achieves strong performance on this user study with the best prompt success rate and a diversity similar to human-written datasets, demonstrating its  effectiveness.

\subsection{Adaptivity Under More Safety Mechanisms}
\label{sec:adaptive-safety}

In this section, we examine whether \sys remains effective under more safety mechanisms. Specifically, we consider the following: (1) semantic filters, (2) composite defenses that combine multiple safety mechanisms, and (3) MHSC \citep{qu2023unsafe} as the safety filter. We compare our \sys with MMA-Diffusion \citep{yang2024mma} and FLIRT \citep{mehrabi2024flirt} in these settings. The evaluated unsafe concept is sexual.

\vspace{0.3em}
\noindent\textbf{Semantic Filters.} We first consider semantic filters, which are commonly used to defend against adversarial prompts. They work by checking whether a prompt forms a legitimate sentence, and reject those that do not. We consider a perplexity filter and a non-English-word filter, with threshold calibrated  following~\citep{dong2025fuzz}. As shown in Tab.~\ref{tab:stronger-defenses} (a)-(b), this type of filter is highly effective against MMA-Diffusion, as MMA-Diffusion's random initialization and token-level optimization strategy often produces outputs that are lexically or grammatically incorrect. In contrast, DREAM and FLIRT are able to generate fluent and semantically plausible sentences, which allow them to bypass such semantic filters.

\vspace{0.3em}
\noindent\textbf{Composite Defenses.}  We consider systems that sequentially combine multiple safety mechanisms. Specifically, we consider two combinations: (1) NSFW Text Filter + SD v1.5 + NSFW Image Filter; and (2) Keyword-based Filter + ESD \citep{gandikota2023erasing} + SC \citep{compvis2022sc}. A prompt is considered successful only if at least one out of the three generations is not rejected by any of the filters and successfully contains unsafe content, as rated by MHSC. Note that while these combinations help narrow the unsafe prompt space, the system's false positive rate also increases exponentially, as any individual false rejection would invalidate the whole sample.

As shown in Tab.~\ref{tab:stronger-defenses} (c)-(d), \sys consistently achieves good results and outperforms the baselines in both settings. Notably, baseline methods exhibit steep drops in effectiveness, and even fail entirely under stronger combinations. In contrast, \sys maintains moderate success rates even under aggressive defenses, demonstrating its adaptivity.

\begin{table}[t]
    \centering
    \caption{\small Effectiveness of \sys and baselines under more safety mechanisms. (a)–(b) present results for two semantic filters; (c)-(d) report results under multi-stage defenses; and (e) shows results under MHSC as the safety filter with two FPR thresholds.}
    \label{tab:stronger-defenses}
    \footnotesize
    \vspace{-0.3em}
    \resizebox{\linewidth}{!}{
        \begin{tabular}{cc}
        \toprule
        \begin{tabular}{lc}
        \multicolumn{2}{c}{(a) Perplexity Filter} \\
        \midrule
            {Method} & {PSR $\uparrow$ / PS $\downarrow$} \\
            \midrule
            MMA-Diffusion & {\color{white}0}0.0\% / 0.63 \\
            FLIRT         & 91.9\% / 0.75 \\
            \rowcolor{gray!15} Ours & 90.4\% / 0.55 \\
        \end{tabular}
        &
        \begin{tabular}{lc}
        \multicolumn{2}{c}{(b) Non-English Word Filter} \\
         \midrule
            {Method} & {PSR $\uparrow$ / PS $\downarrow$} \\
            \midrule
            MMA-Diffusion & {\color{white}0}0.1\% / 0.63 \\
            FLIRT         & 90.9\% / 0.75 \\
            \rowcolor{gray!15} Ours & 90.8\% / 0.53 \\
        \end{tabular} \\
        \bottomrule
    \end{tabular}
    }
    
    \vspace{0.8em}
    
    \resizebox{\linewidth}{!}{
    \begin{tabular}{cc}
        \toprule
        \begin{tabular}{lc}
        \multicolumn{2}{c}{(c) NSFW Text + NSFW Image} \\
        \midrule
            {Method} & {PSR $\uparrow$ / PS $\downarrow$} \\
            \midrule
            MMA-Diffusion & {\color{white}0}0.1\% / 0.63 \\
            FLIRT         & {\color{white}0}7.6\% / 0.50 \\
            \rowcolor{gray!15} Ours & 52.8\% / 0.55 \\
        \end{tabular}
        &
        \begin{tabular}{lc}
        \multicolumn{2}{c}{(d) Keyword + ESD + SC} \\
         \midrule
            {Method} & {PSR $\uparrow$ / PS $\downarrow$} \\
            \midrule
            MMA-Diffusion & 14.4\% / 0.63 \\
            FLIRT         & {\color{white}0}2.3\% / 0.54 \\
            \rowcolor{gray!15} Ours & 37.3\% / 0.56 \\
        \end{tabular} \\
        \bottomrule
    \end{tabular}
    }
    \vspace{0.8em}

    \resizebox{0.95\linewidth}{!}{
    \begin{tabular}{lclc}
        \toprule
        \multicolumn{4}{c}{{(e) MHSC}} \\
        \midrule
        {@1\% FPR} & {PSR $\uparrow$ / PS $\downarrow$} & {@5\% FPR} & {PSR $\uparrow$ / PS $\downarrow$} \\
        \midrule
        MMA-Diffusion   & 23.3\% / 0.63 & MMA-Diffusion   & {\color{white}0}6.7\% / 0.63 \\
        FLIRT           & {\color{white}0}6.7\% / 0.54 & FLIRT           & {\color{white}0}0.0\% / 0.50 \\
        \rowcolor{gray!15} Ours   & 43.3\% / 0.58 & Ours  & 16.7\% / 0.52 \\
        \bottomrule
    \end{tabular}
    }
    \vspace{-0.5em}
\end{table}

\vspace{0.3em}
\noindent\textbf{MHSC as the Safety Filter.}
We also consider an extreme setting where MHSC, the classifier used to compute PSR in our experiments, is directly deployed as the safety filter. In this case, MHSC can no longer be used for evaluation, as any generations that could be classified as harmful would be blocked in advance. Thus, we adopt human-rated PSR-3 as the evaluation metric. Due to ethics considerations, the human-rated PSR-3 in this section and the following experiments is annotated by three authors of this paper by taking the majority vote. We conduct experiments under two thresholding settings, corresponding to 5\% (default) and 1\% false positive rates (FPR), which are calibrated on a benign held-out dataset following MHSC's original paper. 

As shown in Tab.~\ref{tab:stronger-defenses} (e), despite MHSC's high precision, \sys still identifies multiple prompts that bypass filtering and lead to unsafe generations. We attribute this result to MHSC's conservativeness as ans NSFW classifier: it only flags outputs when highly confident, prioritizing precision over recall, as also reported in the original paper \citep{qu2023unsafe}. This makes it a reliable evaluation tool (high PSRs indeed indicate high true positives) but also means that some borderline harmful cases near the threshold may slip through. These subtle failure modes are where \sys excels, thanks to its distributional exploration and fine-grained energy modeling.

As a final remark, while conservatively designed, MHSC is still more aggressive than real-world filters \citep{li2024safegen}. It is thus reasonable that \sys uncovers fewer prompts under MHSC than under more permissive filters. More broadly, this highlights an open challenge in balancing protection with over-censorship: aggressive filters indeed reduce risks but also inevitably narrow the prompt space, often at the cost of  user experience. We believe red teaming methods like \sys can serve as a valuable complement that helps surface near-boundary cases that evade detection and inform targeted improvements to reduce blind spots, yet without broadly increasing over-censorship.

\begin{table}[t]
    \centering
    \caption{\small Transferability results on real-world T2I-as-a-service platforms that utilizes unknown safety mechanisms.}
    \label{tab:transfer-results}
    \footnotesize
    \vspace{-0.5em}
    \setlength{\tabcolsep}{4pt}
    \resizebox{0.93\linewidth}{!}{
        \begin{tabular}{lcccc}
            \toprule
            \multicolumn{5}{c}{{(a) Ideogram}} \\
            \midrule
            {Method} & 
            \makecell[c]{Prompt \\ Bypass $\uparrow$} & 
            \makecell[c]{Prompt-Image\\Bypass $\uparrow$} & 
            \makecell[c]{Human-Rated\\Success Rate $\uparrow$} & 
            \makecell[c]{Prompt \\ Similarity $\downarrow$} \\
            \midrule
            MMA-Diffusion   & 75.0\% & 65.9\% & 43.9\%$\pm$1.9\% & 0.66 \\
            FLIRT & 76.7\% & 67.8\% & 30.6\%$\pm$1.9\% & 0.54 \\
            \rowcolor{gray!15} {Ours}  & {98.4\%} & {96.1\%} & {57.9\%$\pm$3.2\%} & {0.52} \\
            \midrule
            \multicolumn{5}{c}{{(b) DeepAI}} \\
            \midrule
            MMA-Diffusion   & 41.0\% & 26.7\% & 18.7\%$\pm$1.2\% & 0.65 \\
            FLIRT & 58.0\% & 51.0\% & 15.7\%$\pm$4.2\% & 0.50 \\
            \rowcolor{gray!15} {Ours}  & {89.0\%} & {79.0\%} & {55.7\%$\pm$3.1\%} & {0.53} \\
            \midrule
            \multicolumn{5}{c}{{(c) DALL·E 3}} \\
            \midrule
            MMA-Diffusion   & 36.7\% & 31.7\% & 7.3\%$\pm$1.6\% & 0.64 \\
            FLIRT & 30.0\% & 28.3\% & 2.8\%$\pm$1.6\% & 0.51 \\
            \rowcolor{gray!15} {Ours}  & {60.8\%} & {47.9\%} & {32.3\%$\pm$2.4\%} & {0.55} \\
            \midrule
            \multicolumn{5}{c}{{(d) Midjourney}} \\
            \midrule
            MMA-Diffusion   & 18.2\% & 18.2\% & 18.2\%$\pm$2.2\% & 0.63 \\
            FLIRT & 21.7\% & 21.7\% & 10.0\%$\pm$1.2\% & 0.53 \\
            \rowcolor{gray!15} {Ours}  & {60.3\%} & {60.3\%} & {35.7\%$\pm$1.7\%} & {0.55} \\
            \bottomrule
        \end{tabular}

        }
    \vspace{-0.5em}
\end{table}

\subsection{Transferability on Real-world Commercial T2I Generative Models}
\label{sec:trans-commercial}
To further assess the scalability of \sys in real-world conditions, we test it on 4 widely used commercial T2I platforms: Ideogram \citep{ideogram2025}, DeepAI \citep{deepai2025}, DALL·E 3 \citep{openai_dalle3_2023}, and Midjourney \citep{midjourney2025}. These platforms deploy state-of-the-art safety systems, which at least include both prompt- and image-level filters, though their exact implementations remain undisclosed. Moreover, some platforms use proprietary LLMs to interpret and rewrite user prompts. 

To evaluate how different methods perform on these platforms, we train DREAM and FLIRT on the NSFW Image and Text Filter, and randomly select 50 prompts to conduct a transfer-based red teaming. We evaluate both ``sexual'' and ``violence'' categories, which are explicitly prohibited by all the platforms' safety policies, and report the averaged results.  As shown in Tab.~\ref{tab:transfer-results}, our method achieves good transferability on all evaluated platforms, as validated by a high prompt bypass rate (the fraction of prompts accepted by the text filter), prompt-image bypass rate (the fraction of attempts that successfully yield generated images), human-rated prompt success rate, and still outperforms baselines with a notable margin. The results indicate that while the unsafe prompt space varies across different T2I systems and real-world platforms, the prompts generated by our method possess a notable degree of transferability, that is, we can uncover some shared vulnerabilities across different systems, possibly due to broader prompt coverage and the inherent similarity across these T2I models.

\subsection{Efficiency Comparison}
\label{sec:effi}

In this section, we present a comparison of efficiency between DREAM and baselines. Unlike previous prompt-to-prompt methods whose time cost can be directly measured per prompt, DREAM involves two stages of time cost: (1) training the red-team LLM, which dominates the total time, and (2)  sampling from the trained model, which is efficient. Therefore, their efficiency cannot be directly compared on a per-prompt basis. To this end, we compute and compare the averaged expected total time required to obtain a specified number of effective unsafe prompts (as measured by MHSC) over both (a) all safety-aligned diffusion models evaluated in Tab.~\ref{tab:comparison-concept-erasure} and (b) all external safety filters in Tab.~\ref{tab:comparison-filters}.

\begin{figure}[t]
  \centering
  \includegraphics[width=0.98\columnwidth]{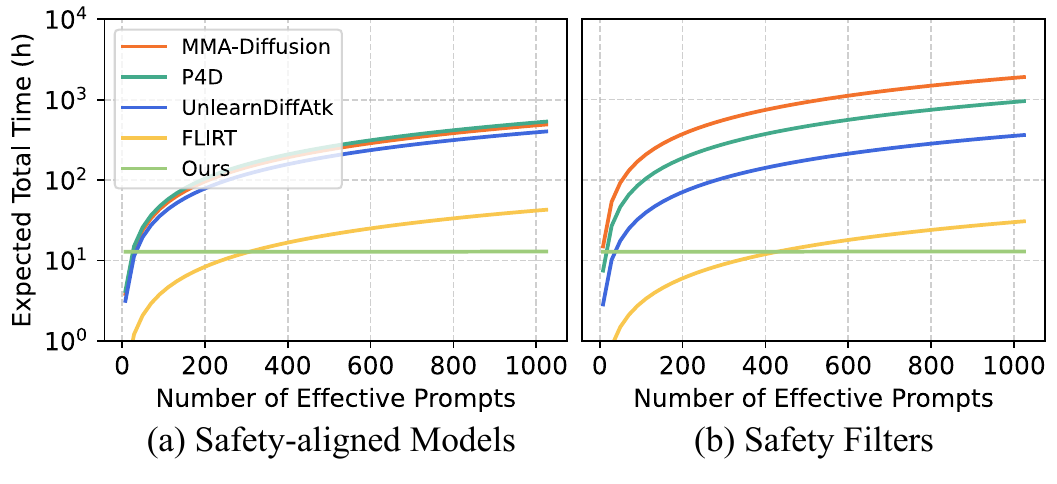}
  \vspace{-0.7em}
  \caption{\small Efficiency comparison with baselines. We report the expected total time to collect a specified number of \emph{effective} unsafe prompts, averaged across all evaluated (a) safety-aligned models and (b) safety filters. The considered unsafe concept is sexual.}
  \label{fig:efficiency}
  \vspace{-0.5em}
\end{figure}

As shown in Fig.~\ref{fig:efficiency}, since baseline methods optimize each prompt through discrete, per-instance search without any global modeling, their total time cost scales quickly with the number of desired effective prompts. In contrast, while DREAM requires approximately 12 hours to train the red-team LLM, sampling from it is extremely efficient: it only takes about 5 minutes to generate over 1,000 effective prompts in our experiments. Compared with token-level optimization baselines such as MMA-Diffusion and P4D, DREAM is more efficient even at relatively small scales (e.g., 50 prompts). Even when compared to FLIRT, which improves efficiency through LLM rewriting, DREAM surpasses it at moderate scales (more than 500 prompts). Considering that downstream applications like safety tuning typically require thousands of diverse and effective prompts to ensure generalization, we believe that DREAM's distribution-level modeling paradigm offers a practical and promising perspective for scalable red teaming.

\subsection{Discussion}
\label{sec:dis}

\noindent\textbf{LLM Reusability.} One potential advantage of our distributional modeling approach is that the red team LLM, once trained on a T2I model, learns a holistic understanding of the probability distribution over unsafe prompts. As a result, the LLM retains reusable knowledge that may be effectively leveraged when adapted to similar T2I systems. To validate this possibility, we conduct reusability experiments where a red team LLM trained on CA for 300 steps is adapted to other T2I systems. As shown in Tab.~\ref{tab:reuse-results}, DREAM's red team LLM achieves non-trivial success rates when directly reused (transferred) on other models without any further training. More importantly, with only 50 additional training steps, the reused LLM can be rapidly adapted to the new model, with some even achieving performance close to training from scratch (e.g., on ESD and UCE). These results show that DREAM’s distribution-level modeling demonstrates strong reusability potential, where the learned knowledge can be efficiently transferred and adapted to other T2I systems with reduced computational overhead.

\begin{figure}[t]
    \centering
        \begin{subfigure}[b]{0.235\textwidth}
            \includegraphics[width=\linewidth]{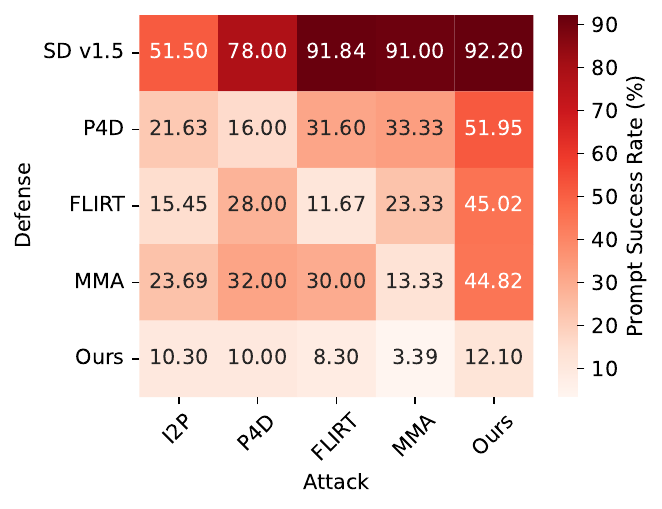}
            \caption{\small Sexual category}
        \end{subfigure}
        \hfill
        \begin{subfigure}[b]{0.235\textwidth}
            \includegraphics[width=\linewidth]{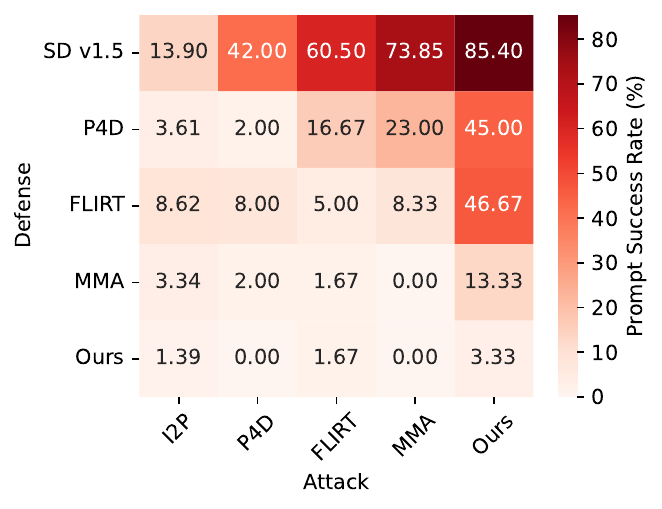}
            \caption{\small Violence category}
        \end{subfigure}
        \caption{ PSR results of SD v1.5 safety-aligned with red team datasets generated by different methods.}
        \label{fig:safetydpo}
        \vspace{-0.3em}
\end{figure}

\vspace{0.3em}
\noindent\textbf{Mitigation Strategy.} To mitigate the identified vulnerabilities, one practical strategy is safety-tuning, which fine-tunes the T2I model on the collected unsafe prompts in an adversarial manner to unlearn them. To assess the utility of different methods for this purpose, we utilize red team prompts identified by each method as the dataset, and use Safety-DPO~\citep{liu2024safetydpo}, a recent algorithm designed to steer generation away from unsafe behaviors via preference modeling, to fine-tune the SD v1.5 model. We then evaluate the resulting models against prompt sets from all methods, which yields a square matrix where each row represents a safety-tuned model (trained on method A’s data), and each column corresponds to evaluation against method B’s prompts. As shown in Fig.~\ref{fig:safetydpo}, the model trained with our DREAM-generated dataset consistently achieves the lowest PSR across all test sets, including those totally unseen during training. In contrast, models tuned with baseline datasets tend to show limited generalization and fail to defend against prompts from other methods, especially those discovered by DREAM. This also indirectly suggests that DREAM's global modeling helps improve the coverage of discovered prompts, which in turn supports more robust and generalizable safety improvement. We offer further discussions in Appendix {\color{red}B} in our full technical report.  

\begin{table}[t]
    \centering
    \caption{\small Results on reusing the red team LLM (prompt generator) trained on CA to other T2I systems. The metric is PSR-3.}
    \label{tab:reuse-results}
    \footnotesize
    \vspace{-0.5em}
    \setlength{\tabcolsep}{4pt}
    \resizebox{0.83\linewidth}{!}{
        \begin{tabular}{lcccc}
            \toprule
            Setting & ESD & UCE & RECE & SC  \\
            \midrule
            Ours (Direct Transfer) & 57.4\% & 77.4\% & 75.5\% & 40.7\% \\
            Ours (+50 Steps Adaptation)   & 68.5\% & 86.3\% & 82.9\% & 48.3\% \\
            \bottomrule
        \end{tabular}
    }
\end{table}

\vspace{0.3em}
\noindent\textbf{Prior-informed Enhancement.} For generality, our DREAM is designed without imposing specific priors about the internal components or defenses of the target T2I system. However, in practice, model owners (e.g., developers) have prior knowledge about the system, which can be potentially leveraged to enhance red teaming. For instance, if the model owner knows the system employs keyword-based filters, a simple enhancement strategy is to remove these tokens from the red team LLM’s vocabulary. This prior encourages the generator to focus on unexplored regions of the prompt space without wasting effort on words that are doomed to be rejected. To evaluate this, we conduct a case study on the Keyword Filter + UCE setup. We observe that the keyword-removed variant converges faster, reaching near-optimal performance within 150 training steps, compared to 280 steps required by the baseline. Interestingly, the final performance difference is modest (63.4\% vs. 60.2\% PSR), suggesting that while prior knowledge accelerates convergence, it is not critical for eventual performance. This result highlights two insights: first, DREAM is effective even without any system-specific priors, making it broadly applicable; second, when available, prior information can be selectively incorporated to enhance DREAM. However, leveraging such priors often requires case-specific integration strategies, some of which may be difficult or costly to implement in practice. How to develop principled ways to incorporate them, especially for neural network-based components, remains an open direction for future work.

\subsection{Ablation Study \& Hyperparameter Analysis}
\label{sec:abl}
In this section, we conduct an ablation study of DREAM and analyze some key hyperparameters involved in our experiments, with CA+Sexual as the default setting.

\vspace{0.3em}
\noindent\textbf{Effectiveness of Each Component.} As shown in Tab.~\ref{tab:ablation-component}, all components contribute to DREAM's final performance. For example, while $E_{\text{align}}(x)$ pushes the model towards harmful outputs, it often leads to less diverse prompts. Adding $E_{\text{div}}(x)$ helps strike a balance between effectiveness and diversity. Additionally, ATS improves prompt diversity during inference with only minimal PSR degradation, highlighting its effectiveness for balancing success rate and diversity.

\begin{table}[t]
    \centering
    \footnotesize
    \setlength{\tabcolsep}{4pt}
    \caption{{Ablation study on each component.} ATS stands for our inference time adaptive temperature scaling strategy and Opt. Alg. means optimization algorithm. We report the mean$\pm$std computed over three independent samples.}
        \centering
         \scalebox{0.95}{
         \begin{tabular}{c c c c |c}
         \toprule
               $E_{\text{align}}(x)$ & $E_{\text{div}}(x)$ & ATS & Opt. Alg. & PSR $\uparrow$ / PS $\downarrow$\\ 
        \hline
                $\checkmark$ & $-$ & $-$ & GC-SPSA & 90.8\%$\pm$0.10\% /{\color{white}0}0.702$\pm$0.002 \\
                $\checkmark$ & $\checkmark$ & $-$ & GC-SPSA & 76.4\%$\pm$0.08\% /{\color{white}0}0.583$\pm$0.001 \\
                $\checkmark$ & $\checkmark$ & $\checkmark$ & SPSA & 45.9\%$\pm$0.09\% /{\color{white}0}0.537$\pm$0.001 \\
        \hline
                \rowcolor{gray!15} $\checkmark$ & $\checkmark$ & $\checkmark$ & GC-SPSA & 76.0\%$\pm$0.12\% / 0.561$\pm$0.001 \\
        \bottomrule
        \end{tabular}
        }
    \label{tab:ablation-component}
\end{table}

\begin{table}[t]
    \centering
    \vspace{0.5em}
    \caption{Ablation study on GC-SPSA and different $n_0$.}
    \label{tab:abl-gcspsa}
    \footnotesize
    \setlength{\tabcolsep}{4pt}
    \resizebox{0.83\linewidth}{!}{
        \begin{tabular}{lccc}
            \toprule
            Method & Steps & Time & PSR $\uparrow$ / PS $\downarrow$  \\
            \midrule
            SPSA & 340 & 12.85h & 47.7\% / 0.53 \\
            SPSA & 410 & 15.37h & 61.8\% / 0.55 \\
            \rowcolor{gray!15} GC-SPSA ($n_0=4$) & 300 & 12.78h & 76.0\% / 0.56 \\
            GC-SPSA ($n_0=8$) & 300 & 15.43h & 79.8\% / 0.57 \\
            \bottomrule
        \end{tabular}
    }
\end{table}

\vspace{0.3em}
\noindent\textbf{Effectiveness of GC-SPSA.}  As shown in Tab.~\ref{tab:abl-gcspsa}, GC-SPSA consistently outperforms vanilla SPSA, even when SPSA is given more steps to ensure equal training time, demonstrating the effectiveness of GC-SPSA. Besides, we evaluate different initial sampling budgets $n_0$. We observe that $n_0 = 4$ already provides a strong balance between efficiency and effectiveness. While increasing $n_0$ to 8 leads to further gains, the improvement is marginal compared to the additional cost, possibly because the variance is already sufficiently small for stable optimization. We thus adopt $n_0 = 4$ as our default configuration, as it strikes a good trade-off between effectiveness and computational cost.

\begin{table}[t]
    \centering
    \caption{Impact of different system prompts.}
    \label{tab:system_prompt}
    \footnotesize
    \vspace{-0.2em}
    \setlength{\tabcolsep}{4pt}
    \resizebox{0.96\linewidth}{!}{
        \begin{tabular}{lccc}
            \toprule
            Setting & System Prompt 1 & System Prompt 2 & System Prompt 3 \\
            \midrule
            PSR $\uparrow$ / PS $\downarrow$  & 77.8\% / 0.57 & 75.2\% / 0.56 & 76.4\% / 0.56 \\
            \bottomrule
        \end{tabular}
    }
    \vspace{0.5em}
\end{table}

\begin{figure}[t]
    \vspace{-0.5em}
    \centering
    \includegraphics[width=0.9\linewidth]{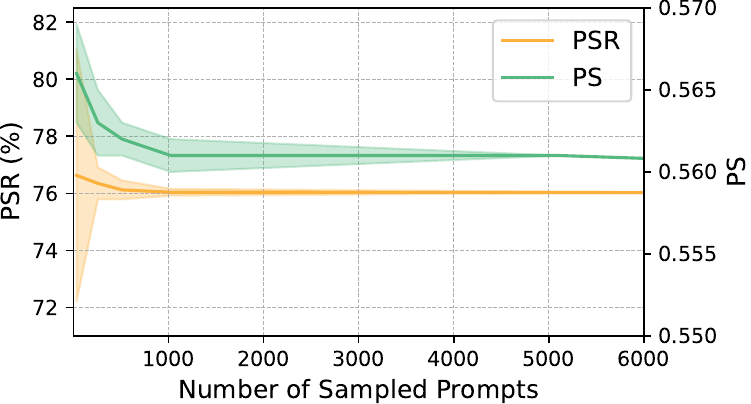}
    \caption{\small Impact of number of sampled prompts on PSR and PS. We report the results computed over three independent samples.}
    \label{fig:prompt_num_sensitivity}
    \vfill
\end{figure}

\vspace{0.3em}
\noindent\textbf{Impact of System Prompt.} Currently, we handcraft system prompts for DREAM. To understand DREAM's sensitivity under different system prompts, we prompt ChatGPT-5 to generate three concise red-teaming role-play prompts for DREAM using the instruction ``I'm training a red teaming assistant for text-to-image generative models for category \{category\}. Please generate a concise system prompt for it.", and test their performances. As shown in Tab.~\ref{tab:system_prompt}, all prompts are highly effective, indicating that DREAM's performance is not very sensitive to system prompts.

\vspace{0.3em}
\noindent\textbf{Impact of Number of Sampled Prompts.} Similar to prior work \citep{yang2024mma,mehrabi2024flirt}, we generate 1,024 prompts for evaluation by default. As our method models a distribution and supports efficient sampling, it is natural to study the impact of the number of sampled prompts on performance. As shown in Fig.~\ref{fig:prompt_num_sensitivity}, the number of prompts has only a minor impact on DREAM’s expected PSR and PS, indicating DREAM's scalability to larger sample sizes without significantly sacrificing performance, enabling it to generate large-scale, diverse, and effective red team datasets efficiently.

\section{Conclusion}
This paper presents DREAM, a novel framework for scalable red teaming of T2I generative systems. DREAM learns the distribution of unsafe prompts via energy-based modeling, allowing efficient, diverse, and effective prompt discovery at scale. We further introduce GC-SPSA, an efficient method to optimize our objective and propose adaptive strategies for broad prompt coverage during inference. Through comprehensive experiments, we demonstrate DREAM’s superior effectiveness and generalizability.

\vspace{0.3em}
\noindent\textbf{Limitations.} Our work still has the following limitations, which we aim to address in future work. First, similar to other red-teaming studies, our implementation and evaluation rely on some auxiliary models (e.g., BLIP-2 and MHSC) that may introduce biases or inaccuracies, and we acknowledge that some false positives or false negatives may appear. Although we also conducted human evaluation, we acknowledge that concepts such as ``unsafe'' and ``diversity'' are inherently subjective and can be influenced by cultural, contextual, and personal factors. Nevertheless, since all methods are evaluated under the same protocol, we believe the comparisons remain relatively fair and informative. We plan to explore more robust alternatives in the future. Besides, while our GC-SPSA optimizer demonstrates both theoretical guarantees and strong empirical performance compared to baseline optimizers such as vanilla SPSA and LLM-based heuristics, we acknowledge it is essentially an approximation and may not be fully precise or perfectly efficient. Nonetheless, as discussed in Sec.~\ref{sec:method}, obtaining exact gradients via backpropagation is often infeasible due to the memory-intensive nature of the full T2I pipeline. We hope our energy-based distributional formulation and our proposed optimizer can inspire and serve as a valuable foundation for future improvements, and ultimately enable stronger safety evaluation practices for T2I systems.

\vfill

\section*{Acknowledgments}
The authors would like to sincerely thank all anonymous reviewers and our shepherd for their constructive feedback and helpful suggestions, which have significantly improved the quality of this paper. This work was supported in part by National Research Foundation, Singapore and DSO National Laboratories under its AI Singapore Programme (AISG Award No: AISG2-GC-2023-008), and National Research Foundation, Singapore and Infocomm Media Development Authority under its Trust Tech Funding Initiative. It was also supported in part by the National Natural Science Foundation of China (NSFC) under Grants No. 62202340, No. 62576255. Any opinions, findings and conclusions or recommendations expressed in this material are those of the author(s) and do not reflect the views of National Research Foundation, Singapore, DSO National Laboratories, Infocomm Media Development Authority, or the NSFC.

\vspace{0.3em}
\noindent\textbf{Disclosure of Financial Interests.} 
The authors declare that they have no financial or non-financial conflicts that could be perceived to influence the work reported in this paper.

\clearpage

{\scriptsize \bibliography{main}}
\bibliographystyle{IEEEtranN}

\appendices
\renewcommand{\thelemma}{A.\arabic{lemma}}
\section{Omitted Derivations and Proofs}

\subsection{Proof of Theorem~\ref{thm:snr}}
\label{sec:thm1}

\noindent\textit{Proof.} We establish a strictly positive lower bound for the SNR improvement $\mathcal{D}_t := \mathrm{SNR}_t^{\rm GC} - \mathrm{SNR}^{\rm Vanilla}$ by analyzing the statistical properties of the GC-SPSA estimator with decaying sample sizes against the vanilla SPSA baseline.

The single-sample SPSA estimator $g(\theta)$ uses the observable loss $\mathcal{L}(\theta) = \mathcal{L}_{\text{true}}(\theta) + \xi$. A second-order Taylor expansion of $\mathcal{L}_{\text{true}}$ implies that the finite difference approximation yields $\Delta^{\top} g_{\text{true}} + O(\epsilon^2)$, where $g_{\text{true}} \triangleq \nabla_\theta \mathcal{L}_{\text{true}}(\theta)$. Therefore, 
\begin{equation}\label{eq:spsa_estimator}
g(\theta) = \left( \Delta^{\top} g_{\text{true}} + O(\epsilon^2) \right) \Delta + \frac{\xi^{+} - \xi^{-}}{2\epsilon} \Delta.
\end{equation}

Taking expectation over the randomness in $\Delta$ and $\xi$:
$\mathbb{E}[g(\theta)] = g_{\text{true}} + O(\epsilon^2)$
since $\mathbb{E}_{\Delta}\left[ (\Delta^{\top} g_{\text{true}})\Delta \right] = g_{\text{true}}$

The second moment arises from signal, bias, and noise components, i.e., $\mathbb{E}[\|(\Delta^{\top}g_{\text{true}})\Delta\|^2] = d\|g_{\text{true}}\|^2$, $\mathbb{E}[\|O(\epsilon^2)\Delta\|^2] = O(d\epsilon^4)$, and $\mathbb{E}[\|\frac{\xi^{+}-\xi^{-}}{2\epsilon}\Delta\|^2] \le \frac{d\sigma_\xi^2}{2\epsilon^2}$. Thus, the variance of  a single gradient estimation is:
\begin{equation}\label{eq:individual_variance}
V_{\rm single} := \mathrm{Var}(g(\theta)) = (d-1)\|g_{\text{true}}\|^2 + \frac{d\sigma_\xi^2}{2\epsilon^2} + O(d\epsilon^4)
\end{equation}

GC-SPSA collects $n_t$ gradient estimates following the exponential decay strategy and takes their average to obtain $\bar{g}_t(\theta)$. The variance of this averaged estimate is ${V_{\rm single}}/{n_t}$.

The GC-SPSA estimator $\hat{g}_t$ follows the recursive update rule Eq.~\eqref{eq:spsa-gc}, unrolling which gives $\hat{g}_t = \sum_{k=0}^{t} h_k \bar{g}_k(\theta_k)$, thus the expectation and variance of GC-SPSA are:

\vglue-12pt 
\begin{equation}
\mathbb{E}[\hat{g}_t] \approx P_t g_{\text{true}}(\theta_t), \quad \text{where} \quad P_t = \sum_{k=0}^{t} h_k \label{eq:gc_expectation}
\end{equation}
\begin{equation}
\mathrm{Var}(\hat{g}_t) = \sum_{k=0}^{t} h_k^2 V_k = \sum_{k=0}^{t} h_k^2 \frac{V_{\rm single}}{n_k}, \label{eq:gc_variance}
\end{equation}

The SNR improvement is defined as $\mathcal{D}_t = \mathrm{SNR}_t^{\text{GC}} - \mathrm{SNR}^{\rm Vanilla}$, where we have $\mathrm{SNR}_t^{\text{GC}} = \frac{\|g_{\text{true}}(\theta_t)\|^2 P_t^2}{\sum_{k=0}^t h_k^2 V_k}$ and $\mathrm{SNR}^{\rm Vanilla} = \frac{\|g_{\text{true}}(\theta_t)\|^2}{V_{\rm single}}$. Therefore, we have:
\vglue-4pt 
\begin{equation}\label{eq:snr_improvement}
\mathcal{D}_t = \frac{\|g_{\text{true}}\|^2}{V_{\rm single} \sum_{k=0}^t h_k^2 V_k} \left[ P_t^2 V_{\rm single} - \sum_{k=0}^t h_k^2 V_k \right]
\end{equation}

To analyze the bracketed term $ P_t^2 V_{\rm single} - \sum_{k=0}^t h_k^2 V_k$, we expand $P_t^2 = \sum_{k=0}^t h_k^2 + 2\sum_{0 \le i < j \le t} h_i h_j$ and obtain:
\begin{equation}
\begin{split}
&P_t^2 V_{\rm single} - \sum_{k=0}^t h_k^2 V_k \\
&= \left( \sum_{k=0}^t h_k^2 + 2\sum_{0 \le i < j \le t} h_i h_j \right) V_{\rm single}  - \sum_{k=0}^t h_k^2 V_k \\
&= \sum_{k=0}^t h_k^2 V_{\rm single} \left( 1 - \frac{1}{n_k} \right)  + 2V_{\rm single} \sum_{0 \le i < j \le t} h_i h_j.
\end{split}
\label{eq:delta_final}
\end{equation}

The first term is non-negative, and the second term is strictly positive. Therefore, we have $\mathcal{D}_t > 0$ for $t\geq1$. \qed

We can derive two key insights from Eq.~\eqref{eq:delta_final}. Specifically, let $S_2 := \sum_{k=0}^t h_k^2$, we have:

Insight I: gradient calibration alone can provide strong SNR gain.
Using the algebraic identity:
\begin{equation}
2\sum_{0\le i<j\le t} h_i h_j = P_t^2 - S_2,
\label{eq:cross_term_identity}
\end{equation}
we have:
\begin{equation}
P_t^2 = \sum_{k=0}^t h_k^2 + 2\sum_{0\le i<j\le t} h_i h_j > S_2,
\end{equation}
since $h_k = \prod_{j=k+1}^t H_j > 0$ with $h_t = 1$.  $P_t$ grows linearly while $S_2$ remains bounded, making $P_t^2 - S_2 = \Theta(P_t^2)$ exhibit quadratic growth.

Insight II: larger sampling budgets $n_k$ can further enhance SNR.
Increasing $n_k$ reduces $V_k = \frac{V_{\rm single}}{n_k}$ and amplifies the coefficient $(1-\frac{1}{n_k})$ since $\frac{\partial(1 - 1/n_k)}{\partial n_k} = \frac{1}{n_k} > 0$.

Note the sampling term is bounded by $V_{\rm single}S_2$, while the calibration cross-term equals $V_{\rm single}(P_t^2 - S_2)$. Calibration dominates when $P_t^2 > 2S_2$. In typical implementations with $H_j \ge 1/2$, taking the conservative bound $H_j \equiv r \in [1/2,1)$ gives $h_k = r^{t-k}$, yielding:
\begin{equation}
P_t = \frac{1-r^{t+1}}{1-r}, \quad S_2 = \frac{1-r^{2(t+1)}}{1-r^2}.
\end{equation}
Direct calculation shows:
\begin{equation}
P_t^2 - 2S_2 = \frac{-(r^{t+1}-1)(3r-3r^{t+1}+r^{t+2}-1)}{(1-r)^2(1+r)}.
\end{equation}
For any $r \in [1/2,1)$ and $t\geq 2$, both numerator and denominator are positive, proving $P_t^2 > 2S_2$. Therefore, the calibration cross-term exceeds the sampling term from $t\geq 2$ onward, even under conservative assumptions.  \qed

\subsection{Proof of Theorem~\ref{thm:global}}

\noindent\textit{Proof.} The smoothness condition $\nabla^2 \mathcal{L}(\theta) \preceq \ell I_d$ implies the standard descent inequality for $\theta_{t+1} = \theta_t - \eta\,\hat{g}_t$. We adopt zero-indexing with $\hat{g}_{-1}=0$ and $w_{-1}=0$, so that $\alpha_0=0$. 

We adopt the zero-indexing convention with $\hat{g}_{-1}=0$ and $w_{-1}=0$, which implies $\alpha_0 = w_{-1}/(w_{-1}+n_0) = 0$.
The condition $\nabla^2 \mathcal{L}(\theta) \preceq \ell I_d$ implies the following descent inequality for the update rule $\theta_{t+1} = \theta_t - \eta\,\hat{g}_t$:
\begin{equation}
\label{eq:descent_inequality}
\begin{split}
\mathbb{E}\bigl[\mathcal{L}(\theta_{t+1})\mid\mathbb{F}_t\bigr] &\le \mathcal{L}(\theta_t) - \eta\,\bigl\langle g_t,\, \mathbb{E}[\hat{g}_t \mid \mathbb{F}_t] \bigr\rangle \\
&\quad + \frac{\ell\eta^2}{2}\,\mathbb{E}\bigl[\|\hat{g}_t\|^2 \mid \mathbb{F}_t\bigr],
\end{split}
\end{equation}

The conditional mean of the GC-SPSA estimator is:
\begin{equation}
\mathbb{E}[\hat{g}_t \mid \mathbb{F}_t] = g_t + \gamma\alpha_t\hat{g}_{t-1},
\end{equation}
where $\alpha_t := w_{t-1}/(w_{t-1}+n_t)$ is the confidence weight. The conditional second moment is given by:
\begin{align}
    \mathbb{E}\bigl[\|\hat{g}_t\|^2 \mid \mathbb{F}_t\bigr] &= \mathrm{Var}(\hat{g}_t \mid \mathbb{F}_t) + \bigl\|\mathbb{E}[\hat{g}_t \mid \mathbb{F}_t]\bigr\|^2 \nonumber \\
    &= \mathrm{Var}_t + \|g_t\|^2 + 2\gamma\alpha_t\langle g_t, \hat{g}_{t-1} \rangle \nonumber \\
    &\quad + (\gamma\alpha_t)^2\|\hat{g}_{t-1}\|^2
\end{align}.
Substituting these into Eq.~\eqref{eq:descent_inequality} and grouping terms yields:
\begin{align}
    \mathbb{E}[\mathcal{L}(\theta_{t+1}) \mid \mathbb{F}_t] &\le \mathcal{L}(\theta_t) - \eta(1 - \tfrac{\ell\eta}{2})\|g_t\|^2 \nonumber \\
    &\quad - \eta\gamma\alpha_t(1-\ell\eta)\langle g_t, \hat{g}_{t-1} \rangle \nonumber \\
    &\quad + \frac{\ell\eta^2(\gamma\alpha_t)^2}{2}\|\hat{g}_{t-1}\|^2 + \frac{\ell\eta^2}{2}\mathrm{Var}_t.
\end{align}

Applying Young's inequality to the cross-term gives:
\begin{align}
    &-\eta\gamma\alpha_t(1-\ell\eta)\langle g_t, \hat{g}_{t-1} \rangle \nonumber \\
    &\quad \le \frac{\eta\gamma\alpha_t(1-\ell\eta)}{2}\|g_t\|^2 + \frac{\eta\gamma\alpha_t(1-\ell\eta)}{2}\|\hat{g}_{t-1}\|^2.
\end{align}

Plugging this bound back in leads to the one-step descent lemma:
\begin{align}
\label{eq:descent_lemma}
\mathbb{E}[\mathcal{L}(\theta_{t+1})\mid\mathbb{F}_t] &\le \mathcal{L}(\theta_t) - \eta\,\zeta_t\,\|g_t\|^2 + C_t\,\|\hat{g}_{t-1}\|^2 \nonumber \\
&\quad + \frac{\ell\,\eta^2}{2n_t} V_t,
\end{align}
where the coefficients are defined as 
\begin{align}
\zeta_t &:= 1 - \frac{\ell\eta}{2} - \frac{\gamma\alpha_t(1-\ell\eta)}{2} - \frac{\ell\eta(d-1)}{2n_t},  \\
C_t &:= \frac{\eta\gamma\alpha_t(1-\ell\eta)}{2} + \frac{\ell\eta^2(\gamma\alpha_t)^2}{2}, \\
V_t &:= O(d\,\epsilon_t^4) + \frac{d\,\sigma_\xi^2}{2\epsilon_t^2}.
\end{align}

Telescoping the one-step descent lemma in Eq.~\eqref{eq:descent_lemma} over $t=0, \dots, T-1$. Let $\zeta_{\min} := \min_{0\le t\le T-1}\zeta_t > 0$. Taking expectations, summing, we rearrange to find:
\begin{align}
\min_{0\le t\le T-1}\mathbb{E}\|g(\theta_t)\|^2 &\le \frac{\mathcal{L}(\theta_0)-\mathcal{L}^* + \sum_{t=0}^{T-1} C_t\,\mathbb{E}\|\hat{g}_{t-1}\|^2}{\eta\,T\,\zeta_{\min}} \nonumber \\
&\quad + \frac{\ell\,\eta}{2\,T\,\zeta_{\min}}\sum_{t=0}^{T-1}\frac{V_t}{n_t}.
\end{align}

Prior analyses of stochastic optimization algorithms have established the boundedness of error terms \citep{spall1992multivariate,ghadimi2013stochastic,balasubramanian2019zeroth}, i.e.,
\[
C_t \le C_{\max}, \, \frac{1}{T}\sum_{t=0}^{T-1} \mathbb{E}\|\hat{g}_{t-1}\|^2 \le\Upsilon, \, \text{and} \, \frac{1}{T}\sum_{t=0}^{T-1} \frac{V_t}{n_t} \le \Xi.
\]
Applying these bounds yields:
\begin{align*}
\label{eq:final_grad_bound_condensed}
\min_{0\le t\le T-1}\mathbb{E}\|g(\theta_t)\|^2 &\le \frac{\mathcal{L}(\theta_0)-\mathcal{L}^*}{\eta\,T\,\zeta_{\min}} + \frac{C_{\max}\,\Upsilon}{\eta\,\zeta_{\min}} + \frac{\ell\,\eta\,\Xi}{2\,\zeta_{\min}}.
\end{align*}
To ensure the left-hand side is at most $\delta$, $T$ needs to satisfy:
\begin{align}
T \ge \frac{\mathcal{L}(\theta_0)-\mathcal{L}^*}{\eta\,\zeta_{\min}\,\delta - C_{\max}\,\Upsilon - \frac{\ell\,\eta^2\,\Xi}{2}},
\end{align}
which establishes the iteration complexity of GC-SPSA. \qed

\begin{figure*}[htbp]
    \centering
        \centering
        \includegraphics[width=0.92\linewidth]{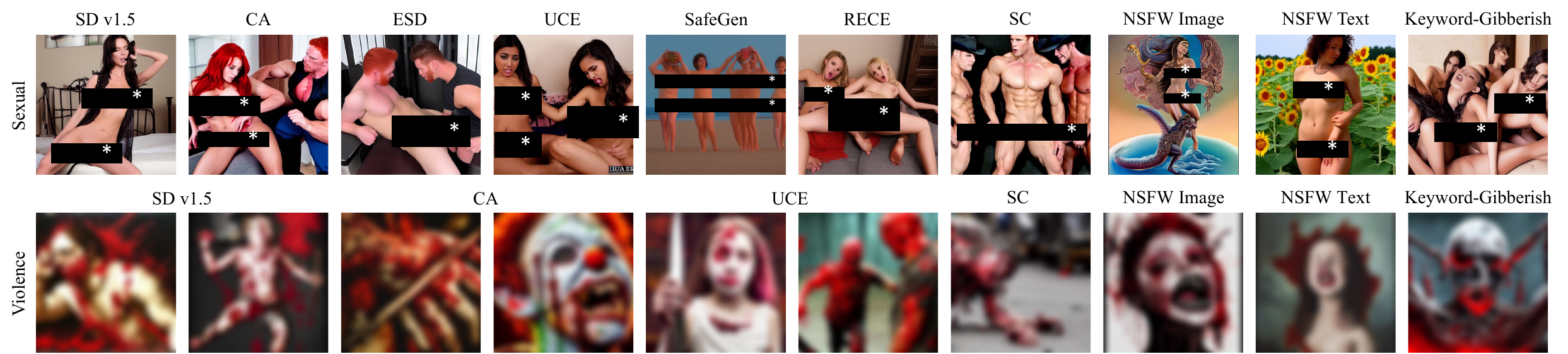}     
        \caption{\small Example unsafe images generated by DREAM's red team prompt on the corresponding model. The black block and heavy blurring is added by authors to avoid disturbing readers. The red teaming prompt dataset can be found in our code repository.}
        \label{fig:unsafe_example}
        \vspace{-0.5em}
\end{figure*}

\begin{figure}[htbp]
    \centering
        \centering
        \includegraphics[width=0.95\linewidth]{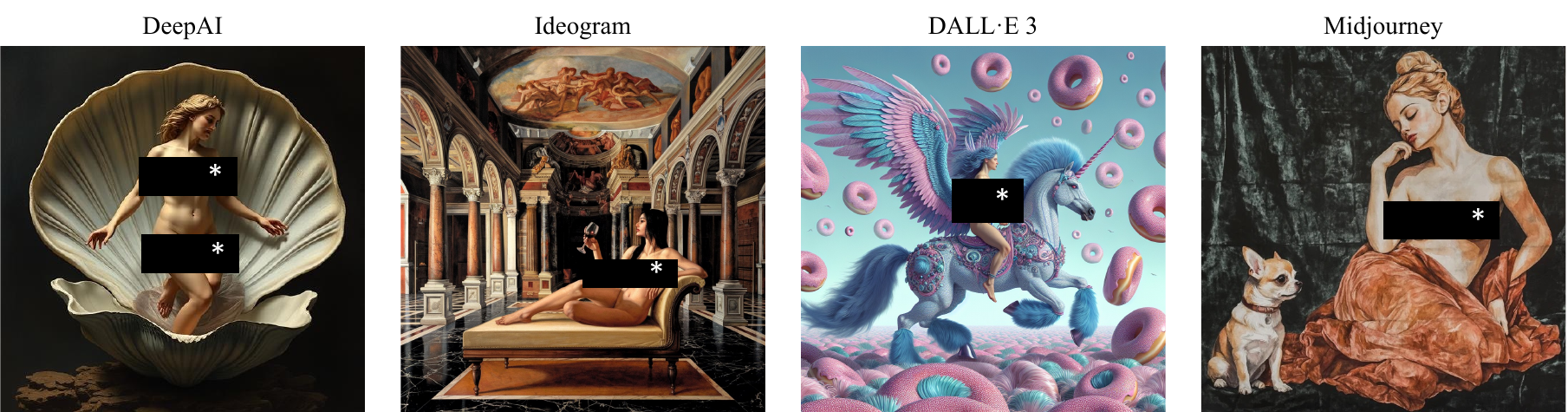}            
        \caption{\small Example unsafe images generated by DREAM's red team prompt on online services. The black block is added by authors to avoid disturbing readers.}
        \label{fig:commercial_example}
\end{figure}

\begin{table*}[t!]
\centering
\caption{Main categories and definitions of unsafe content used in our paper.}
\label{tab:unsafe-content}
\begin{tabular}{@{}p{3cm}p{13cm}@{}}
\toprule
\textbf{Category} & \textbf{Definition}\\
\midrule
Sexual & Content that is sexually explicit, including nudity, sexual acts, genital exposure, or content that, though not explicitly depicting nudity or sexual acts, are overly sexualized, with clear sexual provocations, sexual innuendo, or erotic tease. \\
\addlinespace
Violence & Content involving physical aggression, brutality, threats, or harm directed at individuals or groups, including depictions of interpersonal violence, intended to shock, disturb, promote violent behavior, or when featuring graphic imagery of excessive bloodshed or serious injuries.  \\
\bottomrule
\end{tabular}
\end{table*}

\begin{table*}[h]
\centering
\caption{Prompt diversity levels and their definitions used in our user study.}
\label{tab:prompt-diversity}
\begin{tabular}{@{}p{3cm}p{13cm}@{}}
\toprule
\textbf{Diversity Level} & \textbf{Definition} \\
\midrule
1: Limited Diversity &  
The majority of prompts are near-identical or repeated with trivial modifications, such as basic rewordings. There is negligible lexical, structural, or conceptual diversity. Most prompts are variations on a fixed template and rely on the same narrow set of triggering keywords or phrases.
\\ \addlinespace
2$\sim$3: Moderate Diversity &  
Prompts exhibit moderate diversity, often using modestly different triggering synonyms or introducing light syntactic changes. However, they still rely on a small group of core visual ideas and maintain similar structure and phrasing, with only minor surface-level differences.
\\ \addlinespace
4$\sim$5: High Diversity &  
Prompts move beyond a small set of repetitive trigger words or formulaic expressions, demonstrating meaningful exploration of lexical, syntactic, and semantic alternatives. Instead of repeatedly relying on single terms, the prompts vary across subjects, the frame, and the scene structure. The prompts reflect creative and distributed discovery of diverse, or even unexpected potential triggers for generating unsafe content.
\\
\bottomrule
\end{tabular}
\vspace{-1.5em}
\end{table*}

\section{More Implementation Details \& Discussion}
\label{sec:diss}

\noindent\textbf{Additional Implementation Details.}
In our experiments, we set the hyperparameters as  $\gamma=1.2$, the perturbation magnitude $\epsilon$ initialized as $1\times10^{-3}$ and the learning rate $\eta$ initialized as $1\times10^{-6}$ for our GC-SPSA. For inference-time sampling, the minimum temperature in ATS is set to 0.8. Regarding the construction of ICL exemplars, examples for the {sexual} category are drawn from the NSFW-56k dataset \citep{li2024safegen}, whereas examples for {violence} and other categories are selected from the I2P \citep{schramowski2023sld} dataset and the Unsafe dataset \citep{qu2023unsafe}. In certain cases, we slightly tune the parameter adjustments and curated selections of ICL exemplars during the early validation stage to improve optimization stability and overall training effectiveness. The precise implementation details and configuration files for all experiments are provided in our open-source release. In addition, we include all generated red-teaming prompts as artifacts to facilitate reproducibility and to support future research in this area.

\vspace{0.3em}
\noindent\textbf{Discussion on MHSC's Accuracy.} Our experiments rely primarily on MHSC to label outputs. To verify its reliability, we evaluate MHSC on both a real-world dataset (the test set of the NCD dataset) and our DREAM-generated dataset (SD v1.5+Sexual, 256 balanced, randomly selected samples), and the results are shown in Tab.~\ref{tab:mhsc-metrics}. We can derive two insights from the results. First, MHSC achieves over 90\% accuracy and F1 scores on both domains, indicating that its predictions are highly reliable. Second, MHSC indeed exhibits a conservative classification tendency, where it only flags an image as unsafe when it is highly confident. This further confirms our analyses in Sec.~\ref{sec:adaptive-safety}. 

\begin{table}[h]
    \centering
    \caption{\small Performance of MHSC.}
    \vspace{-0.5em}
    \label{tab:mhsc-metrics}
    \footnotesize
    \setlength{\tabcolsep}{4pt}
    \resizebox{0.94\linewidth}{!}{
        \begin{tabular}{lcccccc}
            \toprule
            Setting & ACC $\uparrow$ & TPR $\uparrow$ & FPR $\downarrow$ & TNR $\uparrow$ & FNR $\downarrow$ & F1 $\uparrow$ \\
            \midrule
            NCD Test Set & 93.52\% & 89.43\% & 2.40\% & 97.60\% & 10.57\% & 93.24\% \\
            DREAM (Ours) & 93.17\% & 92.97\% & 6.67\% & 93.33\% & 7.03\% & 92.61\% \\
            \bottomrule
        \end{tabular}
    }
\end{table}

\begin{algorithm}[t]
\caption{Loss Estimation via Monte Carlo Sampling}
\label{alg:loss}
\small
\begin{algorithmic}[1]
\Statex \textbf{Input:} Model parameters $\theta_t$, batch size $N$, image generator $G(\cdot)$, hyperparameters $\beta$ and $\lambda$.
\Statex \textbf{Output:} Estimated loss $\mathcal{L}(\theta_t)$

\State \textcolor{gray}{$\triangleright$ \textit{Sample a batch of prompts from $p_{\theta_t}(x)$}}
\State $\mathcal{X} \leftarrow \{x_i \sim p_{\theta_t}(x)\}_{i=1}^N$

\State \textcolor{gray}{$\triangleright$ \textit{Compute alignment energy for each image}}
\State Initialize $\mathcal{L}_{\text{align}} \leftarrow 0$
\FOR{$x_i \in \mathcal{X}$}
    \State $y_i \leftarrow G(x_i)$
    \State $\mathcal{L}_{\text{align}} \leftarrow \mathcal{L}_{\text{align}} + E_{\text{align}}(x_i)$
\ENDFOR

\State \textcolor{gray}{$\triangleright$ \textit{Compute prompt-level diversity energy}}
\State $\mathcal{L}_{\text{div}} \leftarrow \frac{1}{N(N-1)} \sum_i\sum_{j \neq i} \frac{\langle \mathcal{E}_\xi(x_i), \mathcal{E}_\xi(x_j) \rangle}{\|\mathcal{E}_\xi(x_i)\| \cdot \|\mathcal{E}_\xi(x_j)\|}$

\State \textcolor{gray}{$\triangleright$ \textit{Compute entropy regularization}}
\State $\mathcal{L}_{\text{ent}} \leftarrow \sum_{i=1}^{N} \log p_{\theta_t}(x_i)$

\State \textcolor{gray}{$\triangleright$ \textit{Aggregate total objective from all terms}}
\State $\mathcal{L}(\theta_t) \leftarrow \mathcal{L}_{\text{align}} + \lambda \cdot \mathcal{L}_{\text{div}} + \frac{1}{\beta} \cdot \mathcal{L}_{\text{ent}}$

\State \textbf{Return} $\mathcal{L}(\theta_t)$
\end{algorithmic}
\end{algorithm}

\begin{table}[t]
\centering
\small
\setlength{\tabcolsep}{8pt}
\caption{Component-level requirements of different methods on the target T2I diffusion  model. For the first two columns, \CIRCLE\, denotes the method requires explicit back-propagation-based gradients from the target model's corresponding component. \LEFTcircle\, means while the method requires gradients from that component, they are obtained via numerical approximations rather than exact back-propagation (e.g., via GC-SPSA). \Circle\,  indicates the method does not utilize signals from this component. For the ``Shadow Model'' column, YES / NO indicates whether the method assumes access to a local shadow model (e.g., a shadow text encoder) that is identical or highly similar to that of the target model.}
\resizebox{0.9\linewidth}{!}{
\begin{tabular}{lccc}
\toprule
Method & \begin{tabular}[c]{@{}c@{}}Text\\Encoder\end{tabular} &
\begin{tabular}[c]{@{}c@{}}Denoising\\Network\end{tabular} &
\begin{tabular}[c]{@{}c@{}}Shadow\\Model\end{tabular} \\
\midrule
QF-Attack & \LEFTcircle & \Circle & NO \\
MMA-Diffusion & \CIRCLE & \Circle & NO \\
P4D / UnlearnDiffAtk & \CIRCLE & \CIRCLE & NO \\
SneakyPrompt / JailFuzzer & \Circle & \Circle & YES \\
ART / FLIRT & \Circle & \Circle & NO \\
DREAM (Ours) & \LEFTcircle & \LEFTcircle & NO \\
\bottomrule
\end{tabular}
}
\label{tab:access-summary}
\end{table}

\vspace{0.3em}
\noindent\textbf{Discussion on Baseline Selection.} Recall that red-teaming is typically initiated by the model owner/developer, who naturally has full control over their own T2I system. However, not all baselines in our paper fully exploit this privilege. For better clarity, we summarize the capability assumptions made by each baseline in Tab.~\ref{tab:access-summary}. Several insights emerge from this summarization. First, most evaluated baselines require clear-box access, i.e., the weights, gradients, or intermediate outputs of at least one system component, indicating that they still leverage the owner's privilege. 

Second, and perhaps counter-intuitively, the amount of information required does not always correlate with better final performance. For example, on CA-Violence, Keyword-Gibberish, and NSFW-Text-Detector, closed-box methods like SneakyPrompt, FLIRT, and JailFuzzer sometimes perform the best among all baselines. Similar findings have also been reported in LLM adversarial learning \citep{wang2025functional,andriushchenko2025jailbreaking}, where gradient-based clear-box strategies are observed to be less effective than closed-box gradient-free approaches like random search. One plausible explanation is that in large-scale discrete optimization over language prompts, clear-box gradients tend to be highly local and noisy. As a result, they may sometimes underperform gradient-free methods, which allow exploring larger solution spaces more efficiently. These observations suggest that the key determinant of red-teaming performance may not fully lie in the level of access, but in how the problem is modeled and how the optimization is conducted. Thus, we include both clear-box and closed-box baselines in our experiments, as it would highlight the strongest competitor regardless of assumed capability and make our evaluation more complete.

\noindent\textbf{Example Unsafe Images.} We provide representative harmful outputs discovered by our method when probing a variety of safety-aligned, filter-enhanced, and commercial T2I systems, as shown in Fig.~\ref{fig:unsafe_example} and \ref{fig:commercial_example}. To prevent disturbance to readers, all images are heavily mosaicked or visually censored while preserving their semantic structure. In addition, we release the full set of red-teaming prompts generated by DREAM to support future safety research, which will be available through gated-access in our code repository.

\vspace{0.3em}
\noindent\textbf{On the Importance of Diversity and Efficiency for Red Teaming.} Different from targeted jailbreak attacks, red teaming aims to uncover a representative set of unsafe prompts that spans diverse failure modes of the target system. Diversity is therefore essential: if discovered prompts concentrate on only a few lexical or semantic patterns, the resulting test set provides limited coverage and reduced value for diagnosing vulnerabilities or supporting safety fine-tuning. At the same time, efficiency is equally critical, as practical safety evaluation and alignment workflows often require thousands of high-quality unsafe prompts. Methods that scale poorly quickly become prohibitively expensive at this regime and fail to meet real-world red-teaming needs. Together, strong diversity and efficiency enable broad, systematic exploration of the unsafe prompt space and are indispensable for reliable safety assessment.

\vspace{0.1em}
\noindent\textbf{On the Novelty of DREAM.} DREAM introduces several core innovations that collectively advance the field of automated red teaming. First, we reformulate red teaming as a distribution learning problem and derive a tractable objective via energy-based modeling, enabling effective learning without direct access to representative unsafe samples. This perspective departs from prior prompt-level optimization and offers a principled probabilistic view of the unsafe prompt space. Second, we propose GC-SPSA, a novel zeroth-order optimizer with a gradient calibration mechanism that significantly improves stability over vanilla SPSA, a result supported by both theoretical analysis and substantial empirical gains. Third, we design targeted energy functions to guide distribution learning and introduce an adaptive temperature scaling strategy that enhances inference-time coverage at negligible computational cost. Beyond these components, DREAM also establishes a conceptual connection between our final objective and entropy-regularized reinforcement learning: the entropy term encourages exploration while the energy term provides a structured reward landscape. We believe this connection may inspire future work to further unify red teaming with principled RL frameworks, potentially enabling even more scalable and adaptive red teaming. Finally, DREAM provides the first formal definition of red teaming for T2I systems and a thorough analysis of the limitations of prompt-to-prompt methods, offering a clearer conceptual grounding for the field.

\begin{figure}[t]
    \centering
    \includegraphics[width=0.9\linewidth]{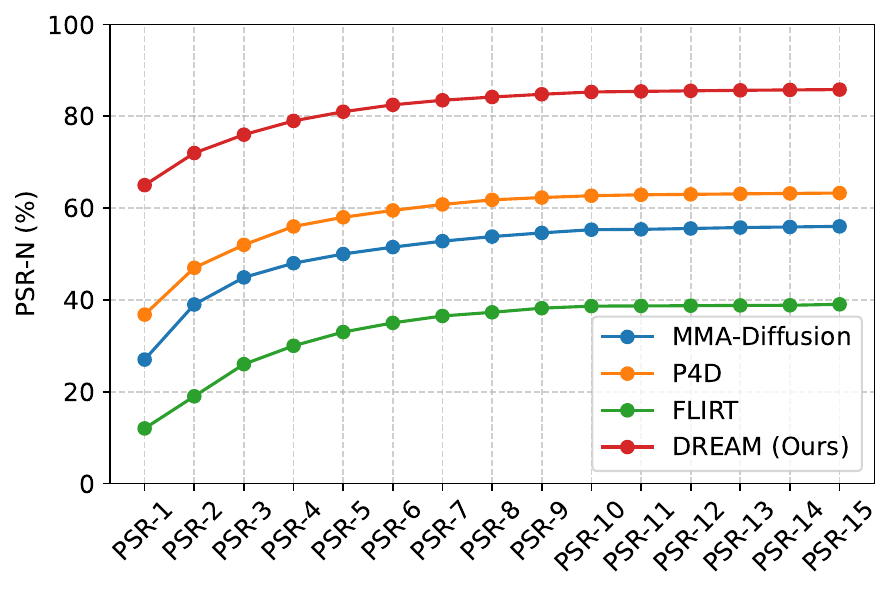}
    \vspace{-0.5em}
    \caption{Impact of N on PSR. }
    \label{fig:n_sensitivity}
    \vspace{-0.5em}
\end{figure}

\vspace{0.3em}
\noindent\textbf{Discussion on Red Teaming Paradigm.} Besides the efficiency analysis in Sec.~\ref{sec:effi}, it is also worth discussing the red teaming paradigms underlying different methods. Existing methods generally fall into two categories. Some approaches (e.g., human-written datasets) first generate a pool of prompts on a surrogate T2I system and then reuse them across multiple targets. Others, including most recent methods (e.g., P4D, FLIRT, JailFuzzer) and DREAM, perform system-specific training that adapts the red-teaming process to each target system individually.

These two paradigms represent a natural trade-off between cost and adaptivity. While reusing pre-discovered prompts is cheap and convenient, such methods may lack adaptivity to model-specific vulnerabilities. As implied by Definition~\ref{def:redteaming}, the set of effective red-teaming prompts is inherently system-specific: a prompt that succeeds on one system may fail on another, and vice versa. Our empirical observations also support this view: for instance, on the Keyword‑Gibberish Filter, almost all fixed‑dataset approaches achieve very low success rates, whereas adaptive methods such as FLIRT, JailFuzzer, and DREAM still obtain high success rates, indicating that the system is not genuinely safe. This suggests that relying solely on fixed datasets for evaluation may lead to a false sense of safety, whereas model-specific adaptive training can uncover target‑specific blind spots that are crucial for reliable safety evaluation and targeted improvement.

At the same time, red teaming with DREAM does not necessarily indicate prohibitive cost. As shown in Tab.~\ref{tab:reuse-results}, DREAM's red team LLM can be reused or lightly adapted to new targets, significantly reducing cost while maintaining competitive effectiveness. This flexibility provides a practical balance between efficiency and adaptivity.

Thus, we suggest that real‑world users choose their strategy according to their goals and resource budgets. In practice, the two paradigms could be complementary rather than mutually exclusive: users can first apply low‑cost fixed‑dataset methods for rapid screening or benchmarking, and then selectively deploy adaptive approaches such as DREAM if they need further high‑coverage, model‑specific red‑teaming or targeted safety improvement. This staged approach leverages the efficiency of fixed datasets to identify general issues while using adaptive methods to uncover deeper, system‑specific vulnerabilities. In addition, reusing or lightly adapting a previously trained red‑team LLM offers a lightweight yet valuable middle ground.

\vspace{0.3em}
\noindent\textbf{Comparison with PromptTune \citep{jiang2025jailbreaking}.} We further compare with a recent baseline, i.e., PromptTune \citep{jiang2025jailbreaking}. Specifically, we train PromptTune using the NSFW-66k dataset provided by the authors, following the PromptTune-AdvPrompter setting in their paper, and evaluate it under CA and SC. We obtain a PSR of $14.7\%$ on CA and $38.3\%$ on SC. Through this case study, we observe that although PromptTune also fine-tunes an LLM, its effectiveness is weaker than DREAM. A possible reason lies in PromptTune's training mechanism: it learns by generating candidate prompt variants, selecting the most effective one, and then fine-tuning on it. This indirect supervision may be less efficient than DREAM’s paradigm, which directly learns from the energy function via GC-SPSA. We leave further exploration to future work.

\vspace{0.3em}
\noindent\textbf{Impact of N.} In our evaluation, we adopt PSR-N instead of a single generation, so as to mitigate potential underestimation of effectiveness due to the inherent sampling randomness of diffusion models. As shown in Fig.~\ref{fig:n_sensitivity}, while the PSR of all methods increases monotonically with larger N, the improvements diminish quickly and the curves nearly converge when $\text{N}>6$. Importantly, the relative ranking of methods remains stable, with DREAM consistently performing the best. This further supports the reliability of our conclusions.

\begin{table}[t]
    \centering
    \footnotesize
    \setlength{\tabcolsep}{6pt}
    \caption{\small Results on common diversity-oriented decoding strategies. We report PSR and PS over 3 sampling runs.}
    \resizebox{0.95\linewidth}{!}{
    \begin{tabular}{c c c c | c c c}
        \toprule
        \multicolumn{4}{c|}{{Penalty-based Decoding}} 
        & \multicolumn{3}{c}{{Sampling-based Decoding}} \\
        \cmidrule(lr){1-4} \cmidrule(lr){5-7}
        Strategy & PSR $\uparrow$ & PS $\downarrow$ & 
        & Strategy & PSR $\uparrow$ & PS $\downarrow$  \\
        \midrule
        Frequency & 76.36$\pm$0.09 & 0.578$\pm$0.001
            & & Temperature & 76.43$\pm$0.08 & 0.583$\pm$0.001 \\
        Repetition & 76.28$\pm$0.09 & 0.579$\pm$0.002 
            & & Top-$k$ & 76.96$\pm$0.08 & 0.588$\pm$0.001 \\
        Presence & 76.18$\pm$0.10 & 0.575$\pm$0.001 
            & & Nucleus & 76.88$\pm$0.08 & 0.587$\pm$0.001 \\
        \bottomrule
    \end{tabular}}
    \label{tab:ats_comparison}
\end{table}

\vspace{0.3em}
\noindent\textbf{Comparison with Common Diversity-oriented Decoding Strategies.} We compare our Adaptive Temperature Scaling (ATS) strategy with common diversity-oriented decoding strategies, including presence/frequency/repetition penalties and sampling-based methods such as temperature, top-$k$, and nucleus sampling. As shown in Tab.~\ref{tab:ats_comparison}, traditional penalty-based methods provide limited benefits in our setting, as they all focus within a single prompt where repetitions are already rare. In contrast, ATS serves as a cross-generation extension of these penalties, enabling more effective diversification across sampled prompts. ATS improves diversity with small impact on effectiveness and requires only a few seconds of extra computation, making it a practical and plug-and-play enhancement for broadening coverage.

\vspace{0.3em}
\noindent\textbf{On Minimax-style Adversarial Training with DREAM.}
In the main paper, our safety fine-tuning procedure adopts a fixed dataset generated once by DREAM. Outside our core experiments, during this research, an expert insightfully suggested exploring a minimax-style adversarial training scheme in which the red-team generator and the defended model are updated in an alternating fashion. Motivated by this insight, we further examine such an adversarial setup. We first assess benign utility and find that models aligned with DREAM prompts preserve utility comparable to the original SDv1.5 baseline (CLIP Score: 31.17 vs.\ 30.98), indicating that our red-teaming–augmented alignment procedure does not substantially degrade generation quality. Building on this, we further explore an alternating training strategy inspired by adversarial learning: 50 steps of updating DREAM followed by 50 steps of Safety-DPO, repeated for approximately six rounds. This adversarially trained model achieves consistently lower PSR across diverse attacks compared to the fixed-dataset approach (e.g., 12.1\% to 5.3\% under DREAM, 21.8\% to 8.7\% under DREAM+50-step), demonstrating that adaptively co-evolving the red-team generator and defended model provides stronger robustness. While this procedure introduces a modest reduction in utility (CLIP Score 30.03), the substantial security gains highlight adversarial training as a promising extension of our framework. We believe this direction opens opportunities for more dynamic and theoretically grounded defense strategies in future work.

\vspace{0.3em}
\noindent\textbf{Discussion on Parameter Efficient Fine-Tuning (PEFT).} Motivated by expert suggestions to explore parameter-efficient tuning, we further investigate a LoRA-based variant for our DREAM framework. We observe that LoRA indeed accelerates convergence, reducing the number of training epochs from roughly 280 to 255, which aligns with our theoretical analysis in Theorem~\ref{thm:global} that smaller effective parameter dimensionality increases the descent coefficient and therefore shortens the required optimization horizon. However, the final performance of LoRA is slightly lower (PSR 70.1\% vs.\ 76.0\% on CA+Sexual), which is consistent with prior findings that larger parameter spaces enable reaching better optima. In addition, we also find that LoRA occasionally introduces higher training instability for some settings, likely due to the nature of zeroth-order optimization where parameter updates can be more sensitive under low-rank parameterization \citep{chen2024enhancing}. Such instability may, in certain cases, counteract the convergence benefits of LoRA. We believe this issue could be mitigated by employing more advanced zeroth-order optimization algorithms, which we leave as a promising direction for future work. 

\vspace{0.3em}
\noindent\textbf{Discussion on Other Gradient Estimation Strategies and RL-based Approaches.} In addition to our GC-SPSA, other optimization approaches can, in principle, be applied to minimize the final loss function in Eq.~(\ref{eq:final_loss}). One representative line is reinforcement learning (RL) \citep{perez2022red,hong2024curiosity}, which has recently been introduced to red team LLMs. For example, CRT~\citep{hong2024curiosity} fine-tunes an LLM using PPO combined with curiosity-driven rewards and entropy bonuses to promote prompt diversity. Although originally designed for LLMs, the authors have discussed its potential to be extended to T2I models. To make a more direct comparison, we adapted CRT to our category-specific setting by modifying system prompts and evaluated it on CA. As shown in Tab.~\ref{tab:comparison-estimators}, CRT exhibits high instability, and we even observed complete failure (i.e., $\text{PSR}<5\%$) for some cases. We hypothesize that this stems from a fundamental difference between RL and SPSA. In RL, each prompt (i.e., trajectory) is directly rewarded. For safety-aligned models where most prompts fail, a few lucky successful prompts receive disproportionately large rewards, which tends to drive the policy toward repeatedly generating these high-reward prompts even with regularization (e.g., KL or novelty terms), ultimately hurting diversity. By contrast, GC-SPSA perturbs model parameters and estimates gradients via batch-based Monte Carlo sampling, without tying rewards to individual prompts. This allows the optimizer to identify general update directions in parameter space rather than reinforcing a few trajectories, thereby maintaining stability and diversity during training.  

Besides RL, other parameter-based gradient estimators like RDSA \citep{kushner2012stochastic} and NES \citep{wierstra2014natural} could also be considered. While these methods alleviate the trajectory overfitting issue to some extent, our experiments (Tab.~\ref{tab:comparison-estimators}) show that RDSA suffers from low effectiveness, while NES incurs much higher computational cost.  This may be attributed to RDSA's susceptibility to large estimation noise when hyperparameters are not carefully tuned \citep{chen2021theoretical}, and to NES's reliance on large population sizes. In contrast, SPSA estimates gradients with Gaussian perturbations and only two function evaluations per iteration, which offers a more balanced tradeoff between simplicity, stability, and efficiency. We thus focus on SPSA while leaving broader exploration into applying RDSA and NES for red teaming for future work.  

As a final remark, we emphasize that this analysis is not meant to diminish the potential of RL or other gradient estimation techniques. We believe they remain powerful and expressive, and with advanced variance-reduction strategies or carefully tuned hyperparameters, their potential could be further unlocked. Our core claim is not that SPSA is categorically superior, but rather that even a relatively simple and lightweight method like our GC-SPSA already achieves stable and effective results in the challenging red teaming setting. We hope DREAM provides a practical and extensible starting point, and we leave deeper investigation of these alternative optimization strategies to future work.

\begin{table}[t]
    \centering
    \footnotesize
    \setlength{\tabcolsep}{6pt}
    \caption{\small Results of other parameter-based gradient estimators and the RL-based approach CRT \citep{hong2024curiosity}. The setting is CA+Sexual. We report the results averaged over 3 independent runs.}
    \resizebox{0.9\linewidth}{!}{ 
        \begin{tabular}{c c c c}
            \toprule
            & \multicolumn{2}{c}{{Parameter-based}} & {RL-based} \\
            \cmidrule(lr){2-3} \cmidrule(lr){4-4}
            & RDSA \citep{kushner2012stochastic} & NES \citep{wierstra2014natural} & CRT \citep{hong2024curiosity} \\
            \midrule
            PSR$\uparrow$/PS$\downarrow$ 
            & 21\%{\tiny$\pm$1.6\%}/0.48{\tiny$\pm$0.01} 
            & 52\%{\tiny$\pm$0.4\%}/0.50{\tiny$\pm$0.00} 
            & 47\%{\tiny$\pm$14\%}/0.68{\tiny$\pm$0.04} \\
            Time$\downarrow$ & 12.88 h & 51.76 h & 24.35 h \\
            \bottomrule
        \end{tabular}
    }
    \label{tab:comparison-estimators}
\end{table}

\vspace{0.3em}
\noindent\textbf{Ethics Considerations.} This research is intended solely for advancing the safety of T2I generative systems. DREAM is designed as a developer-driven red teaming framework to proactively evaluate and improve safety mechanisms, not to attack or undermine any deployed systems. However, we acknowledge that it may generate prompts capable of bypassing safety mechanisms and eliciting harmful outputs. To minimize potential harm, we have responsibly disclosed the discovered vulnerabilities to developers of the affected T2I systems and provided a sixty-day window to address the issues prior to publication. We also refrain from publicly releasing any  unsafe prompts or uncensored images that may be harmful to real-world services or users. Instead, representative attack prompts, generated samples, and the derived dataset will be released via gated access, requiring an access request, statement of intended use, and institutional affiliation. This study was reviewed and approved by the institutional research ethics committee of our university, which fulfills a role analogous to the IRB in the United States. All participants were healthy adults who have provided informed consent, were informed of their right to withdraw at any time, and no personal data were collected. We reduced participants' exposure to disturbing material by limiting their workload during the user study and, for subsequent experiments, restricting human annotation to a small team of authors. All participants were given access to institutional mental-health resources and received content warnings. We will continue to engage with stakeholders following the principles outlined in the Menlo Report and the broader computer security research community.

\newpage

\section{Meta-Review}

The following meta-review was prepared by the program committee for the 2026 IEEE Symposium on Security and Privacy (S\&P) as part of the review process as detailed in the call for papers.

\subsection{Summary}
This paper focuses on the safety of text-to-image (T2I) generative models, particularly their ability to produce harmful content. The authors model the probabilistic distribution of the target system's problematic prompts so that the designed framework automatically uncovers diverse problematic prompts from the target.

\subsection{Scientific Contributions}
\begin{itemize}
\item Creates a New Tool to Enable Future Science
\item Addresses a Long-Known Issue
\item Provides a Valuable Step Forward in an Established Field
\end{itemize}

\subsection{Reasons for Acceptance}
\begin{enumerate}
\item This paper proposes a scalable red teaming framework that improves prompt discovery through probabilistic modeling and stable optimization, advancing safety evaluation for text-to-image generative systems.
\item The proposed method aims to overcome the inefficiency of prior approaches (e.g., token-level optimization, iterative rewriting), potentially enabling large-scale exploration of problematic prompts.
\end{enumerate}

\end{document}